\documentclass[superscriptaddress,twocolumn,showpacs,
amssymb,amsmath,nobibnotes,aps,prd,
nofootinbib]{revtex4-1}
\pdfoutput=1
\usepackage{graphicx,subfigure,bm,color,psfrag,hyperref}
\usepackage{amsfonts}
\usepackage{lipsum}
\usepackage{mathtools}
\usepackage{verbatim}
\usepackage[normalem]{ulem}
\usepackage[dvipsnames]{xcolor}
\hypersetup{colorlinks,linkcolor={blue},citecolor={red},urlcolor={magenta}}

\begin{document}

\title{Is Dark Energy Changing? Probing the Universe's Expansion with present and future astronomical probes}

\author{Mehdi Rezaei}
\email{rezaei@irimo.ir}
\affiliation{Iran Meteorological Organization, Hamedan Research Center for Applied Meteorology, Hamedan, Iran}

\author{Supriya Pan}
\email{supriya.maths@presiuniv.ac.in}
\affiliation{Department of Mathematics, Presidency University, 86/1 College Street, Kolkata 700073, India}
\affiliation{Department of Mathematics, Faculty of Applied Sciences \& Institute of Systems Science, Durban University of Technology, Durban 4000, Republic of South Africa}

\author{Weiqiang Yang}
\email{d11102004@163.com}
\affiliation{Department of Physics, Liaoning Normal University, Dalian, 116029, P. R. China}

\author{David F. Mota}
\email{d.f.mota@astro.uio.no}
\affiliation{Institute of Theoretical Astrophysics, University of Oslo, P.O. Box 1029 Blindern, N-0315 Oslo, Norway}

\begin{abstract}

This study explores the possibility of a time-varying dark energy (DE) equation of state (EoS) deviating from $-1$. We employ a comprehensive dataset of the usual astronomical probes (Type Ia supernovae, baryon acoustic oscillations, Big Bang nucleosynthesis, Hubble data, and Planck 2018 cosmic microwave background (CMB) alongside future mock gravitational wave (GW) distance measurements from the Einstein Telescope. We utilize the Pad\'{e} approximation, a versatile framework encompassing well-known DE models such as constant EoS, Chevallier-Polarski-Linder parameterization and other time-evolving DE parameterizations. Within the Pad\'{e} parameterization, we examine three  
specific forms (Pad\'{e}-I, SPad\'{e}-I, Pad\'{e}-II) applied to both spatially flat and nonflat universes. Pad\'{e}-II exhibits particularly interesting features in terms of the evidence of dynamical DE at many standard deviations. Our results can be summarized as follows. 
Flat universe: When analyzing the combined dataset of standard probes (including CMB) with Pad\'{e}-II in a flat universe, we find a strong preference ($6.4\sigma$) for a dynamical (time-varying) DE EoS. This preference remains significant ($4.7\sigma$) even when incorporating future GW data.
Nonflat universe: in a nonflat universe, the combined standard datasets (without or with CMB) also indicate dynamical DE EoS at a high confidence level ($6.2\sigma$ and $6.4\sigma$, respectively). The addition of GW data slightly reduces the evidence ($3.8\sigma$ and $5.1\sigma$, respectively), but the preference persists.
These results collectively suggest a robust case for dynamical DE in the dark sector. While a nonflat universe is not strongly favored, Pad\'{e}-II hints at a possible closed universe when CMB data are included (with or without GW data).

\end{abstract}

\maketitle

\section{Introduction}
\label{sec-introduction}

Dark matter (DM) and dark energy (DE) are the two main ingredients of the Universe which according to the most up-to-date observational evidence make up nearly 96\% of its total energy density~\cite{Planck:2018vyg}. The  DM sector is responsible for the observed structure formation of our Universe and DE is responsible for its accelerating expansion in the present time. 
The observational evidence further indicates that the $\Lambda$-Cold Dark Matter ($\Lambda$CDM) cosmological model in which DM is pressureless or cold and
$\Lambda$ -- Einstein's positive cosmological constant -- is plugged into the gravitational field equations described by general relativity and acts as the source of DE with an equation of state (EoS), $w=-1$, is a promising cosmological model for modeling the present dynamical universe.  Despite its tremendous successes, $\Lambda$CDM cosmology has faced several challenges both theoretically and observationally.  The cosmological constant problem is one of the biggest issues in the history of modern cosmology ~\cite{Weinberg:1988cp}.  The emergence of cosmological tensions within the $\Lambda$CDM paradigm is another serious issue that has put a question mark on the revision of the $\Lambda$CDM cosmology \cite{DiValentino:2021izs,Abdalla:2022yfr,Perivolaropoulos:2021jda,Kamionkowski:2022pkx}. However, there is no unique way to revise the $\Lambda$CDM cosmology, and in principle, among the existing theories and models in the literature, we do not find any potential reason to prefer any particular theory or model over the others (see for instance~\cite{Bilic:2001cg,Gorini:2002kf,Bagla:2002yn,Nojiri:2003ft,Li:2004rb,Vikman:2004dc,Guo:2004fq,Nojiri:2005vv,Nojiri:2005jg,Amendola:2006kh,Li:2009mf,Paliathanasis:2014zxa,Salvatelli:2014zta,DiValentino:2015ola,Sharov:2015ifa,vandeBruck:2016hpz,Odintsov:2017qif,Mifsud:2017fsy,VanDeBruck:2017mua,Yang:2018qec,Yang:2018pej,Yang:2019jwn,Li:2019san,Rezaei:2019xwo,DiValentino:2019jae,Li:2020ybr,Pan:2020bur,Odintsov:2020voa,DiValentino:2022rdg,Yang:2023qqz,Kumar:2023bqj,Zhai:2023yny,Giare:2024ytc,Giare:2024smz,Wolf:2024eph}, among others) presenting an incomplete list of references.  The simplest extension  of the $\Lambda$CDM cosmology is designed by considering a constant EoS $w$ other than $-1$ where $w$ is assumed to be a free-to-vary parameter in some interval and it is constrained using the observational datasets. This model is widely known as the $w$CDM model. 
The observational constraints on the $w$CDM cosmology indicate that $w$ should be around $-1$, either in the phantom regime ($w < -1$) or in the quintessence regime ($w> -1$) depending on the underlying data~\cite{Rezaei:2020lfy,Yang:2021flj}.  However, the possibility of dynamical DE cannot be excluded. Following some non-parametric approaches, hints for dynamical evolution in the EoS of DE have been reported in the literature ~\cite{Zhao:2017cud,Zhang:2019jsu,Escamilla:2024olw}, and it has been argued that the dynamical DE  could be a potential choice for alleviating the existing cosmological tensions~\cite{Zhao:2017cud,Zhang:2019jsu,Rezaei:2019xwo,Rezaei:2021qwd,Rezaei:2022bkb}. On the other hand, recent measurements of baryon acoustic oscillations from Dark Energy Spectroscopic Instrument (DESI) in combination with the cosmic microwave background radiation and supernovae type Ia datasets have indicated that DE EoS could be dynamical~\cite{DESI:2024mwx} (also see \cite{Giare:2024gpk,Giare:2024oil}).\footnote{According to Refs. \cite{Efstathiou:2024xcq,Huang:2025som}, this evidence needs to be examined further. }

In the present article we have examined such a possibility considering 
some time-evolving DE EoS parameterizations that originate from the general Pad\'{e} approximation \cite{ASENS_1892_3_9__S3_0,baker_graves-morris_1996}. The use of  the general Pad\'{e} approximation in the cosmological setup has gained significant attention in recent times \cite{Gruber:2013wua,Wei:2013jya,Aviles:2014rma,Liu:2014vda,Zhou:2016nik,Rezaei:2017yyj,Capozziello:2017ddd,Mehrabi:2018oke,Rezaei:2019hvb,Rezaei:2023xkj,Fazzari:2025lzd} since it can recover many well known dynamical DE parameterizations introduced in the past. We refer to a number of dynamical DE parameterizations introduced in the literature~\cite{Efstathiou:1999tm,Chevallier:2000qy, Weller:2001gf,Linder:2002et,Jimenez:2003iv,Wetterich:2004pv,Jassal:2004ej,Gong:2005de,Linder:2005ne,Linder:2005dw,Ichikawa:2006qb,Ichikawa:2006qt,Barboza:2008rh,Ma:2011nc,Sendra:2011pt,Li:2011dr,DeFelice:2012vd,Magana:2014voa,Pan:2017zoh,Yang:2017alx,Yang:2018qmz,Li:2019yem,Pan:2019hac,Li:2020ybr,Yang:2021eud,DiValentino:2021rjj,Yang:2022kho,Escamilla:2023oce,Rezaei:2024vtg,Najafi:2024qzm,Giare:2024gpk,RoyChoudhury:2024wri,Escamilla:2024olw,Giare:2024oil,Wolf:2025jlc,Paliathanasis:2025cuc,Kessler:2025kju,Santos:2025wiv,Lee:2025pzo,RoyChoudhury:2025dhe} where many of them can be recovered from the general Pad\'{e} parameterization~\cite{ASENS_1892_3_9__S3_0,baker_graves-morris_1996}.  In this article, our parameterizations are three specific parametric forms of the general Pad\'{e} parameterization which have not been investigated previously and hence these models are new in the literature. Furthermore, in our parameterizations the DE EoS parameters evolve in time, but at the present epoch all of them mimic the cosmological constant ($w =-1$) and recover the concordance $\Lambda$CDM model.  In order to understand the effects of the spatial geometry of the universe on the DE parameters, we have explored the two possibilities, i.e., the spatial flatness of the universe and its curvature. 
We employ the currently available cosmological probes and then the future Gravitational Waves Standard Sirens (GWSS) dataset from the Einstein Telescope in order to understand (i) whether the currently available cosmological probes hint at any deviation from $w =-1$ and (ii) whether the future GWSS probes from the Einstein Telescope indicate the same possibility.

The article is structured as follows: in section \ref{sec-basic-eqns} we discuss the basic equations of DE and introduce different Pad\'{e} parameterizations. In section \ref{sec-data} we discuss the observational data that have been used to constrain 
all the scenarios. Section~\ref{sec-results} discusses the constraints on the proposed Pad\'{e} parameterizations and their implications. In section~\ref{sec-model-comparison} we perform two model comparison techniques and present the results.  Finally, in section \ref{sec-conclusion} we close this article with a brief summary of all findings.

\section{DE parameterizations}
\label{sec-basic-eqns}

We consider that our Universe is homogeneous and isotropic in the large scale and it is described by the Friedmann-Lema\^{i}tre-Robertson-Walker (FLRW) line element expressed in terms of the comoving coordinates $(t, r, \theta, \phi)$ as

\begin{align}
\label{FLRWk}
d{\rm s}^2 = -dt^2 + a^2 (t) \left[\frac{dr^2}{1-kr^2} + r^2 \left(d\theta^2 + \sin^2 \theta d\phi^2\right) \right]\, ,
\end{align}
where $a(t)$ denotes the scale factor of the Universe,  and $k$ describes three different geometries, namely,
spatially flat ($k =0$),  open ($k =-1$) and closed ($k = +1$). 
We further assume that the gravitation sector of the Universe is described by Einstein's general relativity (GR) and the matter sector of the Universe is minimally coupled to gravity. Additionally, there is no nongravitational interaction present between any two matter 
components. Thus, in the FLRW background, the gravitational field equations can be expressed as

\begin{eqnarray}
H^2 = \frac{8 \pi G}{3} \; \sum_{i} \rho_i - \frac{k}{a(t)^2}, \label{EFE1}\\
2 \dot{H} + 3 H^2  =  -  8 \pi G \; \sum_{i} p_i -  \frac{k}{a(t)^2},\label{EFE2}
\end{eqnarray}
 where a dot above a character denotes the cosmic time derivative; $G$ is Newton's gravitational constant; $H \equiv \dot{a} (t)/a(t)$, is the Hubble rate of the FLRW universe and $\rho_{i}$ and $p_{i}$ are respectively the energy density and pressure of the $i$-th component of the fluid. The components in the matter sector are a  pressure-less matter (cold DM+baryons) and a DE fluid with a barotropic EoS.  There is no interaction between any two components, hence,  
 from the conservation equation for the total fluid $\sum_{i} \dot{\rho}_i + 3 H \; \sum_{i} (p_i + \rho_i) = 0$,  one can write down the conservation equation of the $i$-th fluid as $\dot{\rho}_i + 3 H \; (p_i + \rho_i) = 0$.   Using this conservation equation, one can find the evolution of the individual fluid. For pressureless matter, $\rho_{\rm m} = \rho_{\rm m,0} (a/a_0)^{-3}$ ($\rho_{\rm m,0}$ is the present day value of $\rho_{\rm m}$) where $a_0$ refers to the present value of the scale factor. The evolution of DE follows the expression

\begin{eqnarray}
\rho_{\rm de}= \rho_{\rm de,0} \left(a/a_0 \right)^{-3}\exp\left(-3\int^a_{a_0}\frac{w(a')}{a'}da'\right),
\end{eqnarray}
where $w(a)$ is the barotropic EoS of DE and $\rho_{\rm de,0}$ is the present value of the DE density. 
 Notice that eqn. (\ref{EFE1}) can be expressed as $1 = \Omega_K + \sum_i \Omega_i$, where $\Omega_i = \rho_i/\rho_c$ ($\rho_c = 3H^2/8 \pi G$ is the critical density) is dubbed the density parameter for the $i$-th fluid  and $\Omega_k = -k/(a^2 H^2)$ is the curvature density parameter. Thus, for a spatially flat Universe, $\Omega_{k} = 0$; for an open Universe, $\Omega_k > 0$; and for a closed  Universe, $\Omega_{k} < 0$.  Now, concerning the EoS of DE, $w(a)$ if it is not constant, then a variety of phenomenological parameterizations can be proposed. In this work, we motivate the choice of the DE EoS through a novel parameterization namely, the Pad\'{e} parameterization \cite{ASENS_1892_3_9__S3_0} which recovers many existing parameterizations of DE. Originally, the Pad\'{e} approximation  of any arbitrary  function $\mathcal{F}$ of order $(m, n)$ is given by \cite{ASENS_1892_3_9__S3_0} (see also \cite{baker_graves-morris_1996})

\begin{eqnarray}\label{PADE}
\mathcal{F}(x)=\frac{c_0+ c_1 x+c_2 x^2+...+c_m x^m}{d_0+d_1 x+d_2x^2+...+d_nx^n},
\end{eqnarray}
where $(m,n)$ are positive numbers and the coefficients $c_{i}$'s ($i = 1, 2,3,...m$) and $d_{j}$'s ($j = 1, 2, 3,...n$) are any real numbers with the condition that the denominator of eqn. (\ref{PADE}) never vanishes. 
One can notice from eqn. (\ref{PADE}) that for $d_{\rm j}=0$ where $j \geq 1$,   the standard Taylor  series expansion is recovered. 
Now, since $\mathcal{F}(x)$ in eqn. (\ref{PADE}) refers to a general approximation scheme, one can use this approximation for the cosmological variables evolving with time.  Following this one can parametrize the DE EoS  as follows:

\begin{eqnarray}\label{PADE-w}
w(x)=\frac{c_0+c_1x+c_2x^2+...+c_m x^m}{d_0+d_1x+d_2x^2+...+d_n x^n},
\end{eqnarray}
where $x$ represents any dimensionless cosmological variable (for example, $a/a_0$ or any elementary function of $a/a_0$, such as $\ln (a/a_0)$ or others) and  the denominator of eqn. (\ref{PADE-w}) does not vanish for any choice of $d_{j}$ ($j = 1, 2, 3,...n$). 
With eqn. (\ref{PADE-w}), one can establish a general parameterization scheme for the DE EoS.  One can clearly notice that the well known Chevallier-Polarski-Linder (CPL) parameterization ~\cite{Chevallier:2000qy, Linder:2002et}: $w(a) = w_0 + w_a \times (1- a/a_0)$, in which $w_0$ is the present value of the DE EoS and $w_a$ is another free parameter, is a special case of eqn. (\ref{PADE-w}).\footnote{A recent review of the CPL parameterization exploiting all possible astronomical probes is available~\cite{Giare:2025pzu}. } One can also notice that in terms of the cosmological redshift $z$ ($= -1 + a_0/a$),  i.e. choosing $x = z$, the general Pad\'{e} parameterization of eqn. (\ref{PADE-w}) can recover the well known linear EoS of DE~\cite{Cooray:1999da}: $w(z) = w_0+ w_a \times z$, and the polynomial form of $w(z) = \sum_{i} w_i z^{i}$ \cite{Astier:2000as}. 
Moreover, with a suitable choice of the variable $x$, the general Pad\'{e} approximation (\ref{PADE-w}) can also recover some of the oscillating DE parameterizations~\cite{Pan:2017zoh,Panotopoulos:2018sso,Rezaei:2019roe,Escamilla:2024fzq} and many other parameterizations. In particular, this general Pad\'{e} approximation (\ref{PADE-w}) can recover many DE parameterizations from one parameter to many parameters  with the proper choices of the parameters involved. In this article we shall investigate three particular parameterizations of the DE EoS derived from the general Pad\'{e} parameterization (\ref{PADE-w}). To describe the origin of these simplified parameterizations, we start with the following two parameterizations \cite{Rezaei:2017yyj,Rezaei:2023xkj} derived from (\ref{PADE-w}) considering ($m, n$) $=$ ($1$, $1$):

\begin{eqnarray}
&&w(a)=\frac{c_0+c_{1}(1-a/a_0)}{d_0+d_{1}(1-a/a_0)}, \label{pade1}\\
&&w(a)=\frac{c_0+c_{1}\ln{(a/a_0)}}{d_0+d_{1}\ln{(a/a_0)}}, \label{pade2}
\end{eqnarray}
where $d_0\neq 0$ in both  eqns. (\ref{pade1})  and (\ref{pade2}) and  we have chosen $x= 1-a/a_0$ for (\ref{pade1}) and $x = \ln (a/a_0)$ for (\ref{pade2}). Now, these two equations can be expressed as 

\begin{eqnarray}
&&w(a)=\frac{w_0 +w_{1}(1-a/a_0)}{1+w_{2}(1-a/a_0)}, \label{pade1.1}\\
&&w(a)=\frac{w_0+w_{1}\ln{(a/a_0)}}{1+w_{2}\ln{(a/a_0)}}, \label{pade2.1}
\end{eqnarray}
where $w_0 = c_0/d_0$ refers to the present value of the DE EoS;  $w_1 = c_1/d_0$ and $w_2 =d_1/d_0$ are free parameters. Eqn. (\ref{pade1.1}) can be further simplified by setting $w_2 =0$ as follows \cite{Rezaei:2017yyj,Rezaei:2023xkj}
\begin{equation}\label{simplified-pade1.1}
w(a)=\frac{w_0}{1+w_{2}(1-a)}\;.
\end{equation}
The above three parameterizations, namely, (\ref{pade1.1}), \ref{pade2.1} and (\ref{simplified-pade1.1}) have been discussed previously by~\cite{Rezaei:2017yyj,Rezaei:2023xkj},  rather, we investigate the time evolution of the DE EoS parameterizations which at the present epoch mimics the cosmological constant, i.e. $w_0 =-1$. This is motivated because, on the one hand, the $\Lambda$CDM cosmological model has achieved tremendous success over the last several years, but on the other hand, the possibility of the dynamical DE EoS can resolve many existing debates in the literature. Thus, with this setting, one can realize a time-varying nature in the DE EoS but at the present epoch the resulting scenario will recover the $\Lambda$CDM cosmology.  Therefore, we set $w_0 =-1$ in all the above three parameterizations and finally get 

\begin{align}
& w(a)= - \frac{1}{1+w_{2}(1-a/a_0)} + \frac{w_{1}(1-a/a_0)}{1+w_{2}(1-a/a_0)}\;, \label{pade1.2}\\
& w(a)= - \frac{1}{1+w_{2}(1-a/a_0)}\;, \label{simplified-pade1.2}\\
& w(a)= -\frac{1}{1+w_{2} \ln (a/a_0)} + \frac{w_{1}\ln(a/a_0)}{1+w_{2}\ln(a/a_0)}\;, \label{pade2.2}
\end{align}
which, for the sake of convenience, we label as Pad\'{e}-I, SPad\'{e}-I and Pad\'{e}-II, respectively. Thus, with this consideration, these parameterizations belong to a special set of DE parameterizations that remain dynamical for $a < a_0$ but at the present epoch (i.e. at $a=a_0$) they are nondynamical since $w(a)$ takes the value $-1$.

\section{Observational data}
\label{sec-data}

In this section we describe the observational datasets that we have employed to constrain the proposed Pad\'{e} scenarios.  We first consider the currently available cosmological datasets and then we combine the 1000 luminosity distance measurements of the GWSS matching the expected sensitivity of the Einstein Telescope with the current cosmological probes.

\subsection{Cosmological probes}
\label{sec-usual-cosmological-probes}
In the context of cosmological probes, the list of data sets used in this article is as follows. 

\begin{itemize}

\item {\bf Pantheon sample of Sne Ia:} We take the Pantheon sample of Type Ia Supernovae (SNe Ia) comprising 1048 data points  within the redshift range $0.01 < z < 2.26$ from \cite{Scolnic:2017caz}. We label this dataset as {\bf SN}.

\item {\bf Baryon acoustic oscillations:} Concerning  the baryonic acoustic oscillations (BAO) data we have used the radial component of the anisotropic BAO obtained from the measurements of the power spectrum and bispectrum from the BOSS Data Release 12 galaxy sample from \cite{Gil-Marin:2016wya}, the complete SDSS-III Ly$\alpha$-quasar sample \cite{duMasdesBourboux:2017mrl} and the SDSS-IV extended BOSS DR14 quasar sample from \cite{Gil-Marin:2018cgo}. We label this dataset as {\bf BAO}.

\item {\bf Big Bang Nucleosynthesis:} The baryon density information of  Big Bang nucleosynthesis in terms of $\Omega_bh^2 = 0.022 \pm 0.002$ (at the 95\%  confidence level) from \cite{Burles:2000zk} has been used. We label this data point as {\bf BBN}.

\item {\bf  Hubble data:} We use the data on cosmic chronometers, obtained from the spectroscopic techniques applied to passively–evolving galaxies, i.e., galaxies with old stellar populations and low star formation rates. Here we have used a sample of 37 data points from different references of \cite{Jimenez:2003iv,Simon:2004tf,Stern:2009ep,Zhang:2012mp,Moresco:2012jh,Moresco:2015cya,Moresco:2016mzx,Ratsimbazafy:2017vga}. We label this dataset as {\bf $H (z)$}.

\item {\bf Cosmic microwave background:} We have used the cosmic microwave background (CMB) distance prior from the final Planck 2018 release. These distance priors derived from the base $\Lambda$CDM model can be used to replace the global fitting of full data released by Planck in 2018 for the other DE models~\cite{Chen:2018dbv}.
We label this dataset as {\bf CMB}. 

\end{itemize}

\begin{figure}
\centering
\includegraphics[width=0.55\textwidth]{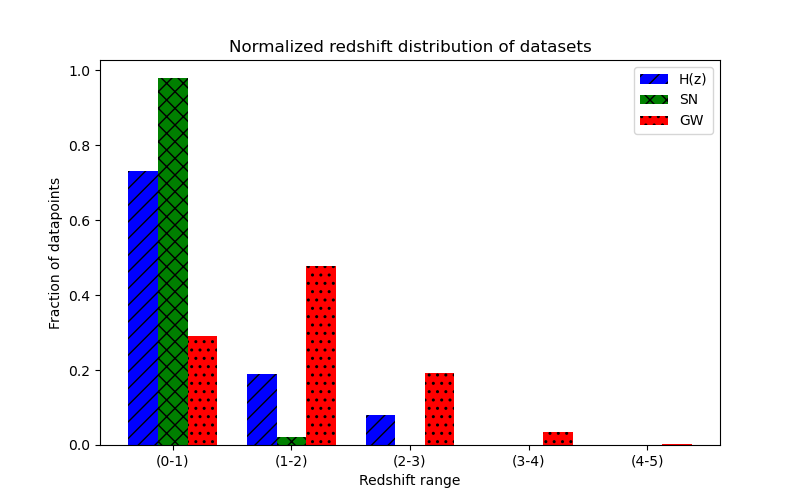}
\caption{Redshift distribution of the SN, $H(z)$ and GW  samples using the normalized histogram description. Note that in the redshift regions $(3--4)$ and $(4--5)$ only the GW data are present,  in particular, only a few data points of the GW sample are in the latter region, and hence the vertical bar in this redshift region is barely visible. }
 \label{figdata}
\end{figure}

\subsection{GWSS luminosity distance measurements}

In this section we describe the  mock GWSS dataset from the Einstein Telescope.  The process describing the extraction of mock GWSS dataset matching the expected sensitivity of the Einstein Telescope is described by \cite{Zhao:2010sz,Cai:2016sby,Wang:2018lun,Du:2018tia,Yang:2019vni,Yang:2019bpr,Yang:2020wby,Matos:2021qne,Pan:2021tpk,Rezaei:2023xkj}, but we describe this procedure briefly here.  The Einstein Telescope is a proposed ground-based third-generation GW detector~\cite{Sathyaprakash:2012jk,Maggiore:2019uih} which is likely to detect $\mathcal{O} (10^3)$ GWSS events after 10 years of full operation. This number of detections is quite optimistic, and it could be fewer in reality~\cite{Maggiore:2019uih}, therefore, the actual picture will be better understood once the exact number of detections after full operation of the Einstein Telescope is known. 
However, following earlier articles in this direction~\cite{Zhao:2010sz,Cai:2016sby,Wang:2018lun,Du:2018tia,Yang:2019vni,Yang:2019bpr,Yang:2020wby,Matos:2021qne,Pan:2021tpk,DAgostino:2022tdk,Rezaei:2023xkj}, we focus on the generation of 1000 mock GWSS in this article and combine this mock dataset with the usual astronomical probes as described in section~\ref{sec-usual-cosmological-probes}. 
To generate this number of mock data points, the identification of the GW sources is crucial. We consider that a GW event is detected from (i) a merger of a Black Hole and a Neutron Star (labeled BHNS) and (ii) a  binary neutron star (BNS) merger. After identifying the possible sources of the GW detection, one needs to find out the merger rate $R(z)$ of the sources. Then using this merger rate, the  redshift distribution of the possible sources, $P (z)$, can be given by ~\cite{Sathyaprakash:2009xt,Zhao:2010sz,Cai:2016sby,Wang:2018lun,Du:2018tia,Yang:2019bpr}

\begin{eqnarray}\label{eqn:P(z)}
    P (z) \propto \frac{4 \pi d_C^2 (z) R (z)}{H (z) (1+z)},
\end{eqnarray}
where $d_C (z) \equiv \int_{0}^{z} H^{-1} (z^\prime) dz^\prime$ stands for the comoving distance, and the merger rate $R(z)$ is given as follows:  $R(z) = 1+2 z$ for $z\leq 1$, $R (z) = \frac{3}{4}(5-z)$  for $1<z<5$, and $R (z) = 0$ for $z > 5$ (see~\cite{Schneider:2000sg,Cutler:2009qv,Zhao:2010sz,Cai:2016sby,Wang:2018lun,Du:2018tia,Yang:2019bpr}).  
Having now $R (z)$ and $P(z)$ from eqn. (\ref{eqn:P(z)}), we sample 1000 sets of 
 ($z_i$, $d_{L} (z_i)$, $\sigma_i$) where $z_i$ denotes the redshift of a possible GW source, $d_L (z_i)$ refers to the measured luminosity distance at redshift $z_i$ and $\sigma_i$ is  the uncertainty of the luminosity distance $d_L (z_i)$. Here  the 1000 mock GWSS luminosity distance measurements have been generated by matching the expected sensitivity of the Einstein Telescope.

In the next step, a fiducial model should be taken because $P (z)$ [eqn. (\ref{eqn:P(z)})] involves the expansion rate $H(z)$ which requires a cosmological model in the background.
A common choice of fiducial model in the literature is $\Lambda$CDM since it represents the simplest cosmological scenario among the existing models, but one can also consider the present Pad\'e parameterizations as the fiducial models. However, in this article, the spatially flat $\Lambda$CDM model has been assumed as the fiducial model and we used the best-fit values of the flat $\Lambda$CDM model parameters as given by Planck 2018~\cite{Planck:2018vyg} to generate the mock GWSS dataset.  
Under the assumption of this fiducial model, one can now determine the luminosity distance at redshift $z_i$ following the relation

\begin{eqnarray}
d_L(z_i) = (1+z_i)\int_0^{z_i}\frac{dz'}{H(z')}\,.
\label{eq:luminosity}
\end{eqnarray}
Lastly, one needs to calculate the uncertainty
linked with the luminosity distance measurements. We consider two types of uncertainties, (i) instrumental uncertainty $\sigma_i^{\rm inst}$, and (ii) weak lensing uncertainty $\sigma_i^{\rm lens}$. The instrumental uncertainty can be approximated as $\sigma_i^{\rm inst} \simeq 2 d_L (z_i)/\mathcal{S}$ ($\mathcal{S}$ denotes  the combined signal-to-noise ratio\footnote{It should be noted that the combined signal-to-noise ratio of the Einstein Telescope should be at least $8$ for a GW detection~\cite{Sathyaprakash:2009xt}.} of the Einstein Telescope) 
\cite{Zhao:2010sz,Cai:2016sby,Wang:2018lun,Du:2018tia,Yang:2019bpr}, and the lensing error is given by $\sigma_i^{\rm lens} \simeq 0.05 z_i d_L (z_i)$~\cite{Zhao:2010sz}.\footnote{For a detailed discussion on the lensing uncertainties, we refer to~ \cite{Hirata:2010ba,Wu:2022vrq}. } Combining the two, the total uncertainty on $d_L (z_i)$ is calculated as  $\sigma_i = \sqrt{(\sigma_i^{\rm inst})^2 + (\sigma_i^{\rm lens})^2}$. With these descriptions, we generated 1000 mock GWSS.

In Fig.~\ref{figdata}, we show the redshift distribution of the SN, $H(z)$ and GW  samples using the normalized histogram description. One can notice that in the redshift region $(4--5)$ only a few GWSS data points are present and hence the vertical bar representing the GW sample is barely visible in this figure.  Note further that other data points, namely, BAO, CMB and BBN are not shown in Fig. \ref{figdata} because these datasets include only a small number of data points compared to the 1000 in the mock GWSS data set, and as a result, the vertical bars representing these data points will not be visible next to the vertical bars for SN, $H(z)$ and GWSS. 

In this article we combine 1000 mock GWSS with the real datasets. 
Usually, GWSS alone is not enough for precise constraints on the cosmological parameters (see for instance \cite{Pan:2021tpk,Matos:2021qne,Li:2023gtu,Chen:2024xkv}),\footnote{We note that, depending on the nature of the model, sometimes this 1000 mock GWSS data set may not be sufficient to properly constrain the model. The present Pad\'{e} parameterizations were not properly constrained by this 1000 mock GWSS set alone. }  but when it is combined with the real datasets such as SN, $H(z)$, BAO, and others (here we combine 1000 mock GWSS with two combined datasets, namely, SN+BAO+BBN+$H(z)$ and SN+BAO+BBN+$H(z)$+CMB), the uncertainties of the cosmological parameters are expected to be improved, and sometimes, depending on the underlying model, the reduction in uncertainties of the model parameters becomes significant~\cite{Wang:2018lun,Du:2018tia,Yang:2019bpr,Yang:2019vni,Pan:2021tpk,Li:2023gtu,Chen:2024xkv}. This indicates that the mock GWSS dataset can increase the precision in estimating some (or all) of the cosmological parameters. However, it should be mentioned that the percentage of precision induced by the mock GWSS depends on many factors, such as the complexities of the underlying cosmological model~\cite{Liao:2017ioi}, and  one may expect that the final picture could eventually alter after  full operation of the Einstein 
Telescope.  Last but not least, we would like to point out that the combination of GWSS and the real datasets may introduce tensions in the cosmological parameters, and such tensions can influence the overall scenario. If the parameters assumed for the fiducial model (here $\Lambda$CDM) differ significantly from those obtained from the combined data sets of the real probes (e.g. SN+BAO+BBN+$H(z)$ and SN+BAO+BBN+$H(z)$+CMB) applied to the Pad\'{e} parameterizations, then this will introduce tensions between the synthetic GWSS data and the combined probes of the real datasets. Because of these tensions, the constraints on the DE parameters will be influenced and the preference for a dynamical DE EoS will be affected as well. In addition to that, these tensions can 
influence the constraining power of the GWSS when it is combined with the real data sets.  As a result of this, how much real constraining power of GWSS is added to the constraints on the Pad\'{e} parameterizations will not be properly assessed.

\begin{table*}
\centering
\caption{Best-fit values of various free and derived parameters at $1\sigma$ ($2\sigma$) confidence level of the four cosmological scenarios in a flat universe considering the combined datasets. }
 \begin{tabular}{c  c  c  c c }
 \hline 
 Model & $\Omega_{m0}$  & $h$   & $w_1$ 
 & $w_2$\\
 \hline 

\multicolumn{5}{c}{SN+BAO+BBN+$H(z)$} \\
 \hline 

 Pad\'{e}-I & $0.261\pm 0.017$  &  $0.6837\pm 0.0097$  & $0.44\pm 0.38$~($0.44\pm 0.77$)& $-0.16\pm 0.37$~($-0.16\pm 0.75$) \\
 
SPad\'{e}-I &  $0.261\pm 0.014$  &  $0.6839\pm 0.0095$  & $-$ & 
 $0.39^{+0.27}_{-0.31}$~($0.39^{+0.50}_{-0.53}$)\\

 Pad\'{e}-II & $0.258\pm 0.017$  &  $0.685\pm 0.010$  &  $0.33\pm 0.23$ ($0.33\pm 0.45$)  &  $-0.70\pm 0.39$~($ -0.70^{+0.82}_{-0.86}$) \\
 
  $\Lambda$CDM &  $0.269\pm 0.009$ &  $0.6857\pm 0.0049$  &  $-$  & $-$ \\

\hline 
\multicolumn{5}{c}{SN+BAO+BBN+$H(z)$+CMB} \\
 \hline 
Pad\'{e}-I & $0.2756^{+0.0085}_{-0.010}$ & $0.6910^{+0.0070}_{-0.0084}$  &  $-0.34^{+0.27}_{-0.24}~(-0.34^{+0.55}_{-0.50})$  &  $0.60^{+0.37}_{-0.23}~(0.60^{+0.71}_{-0.48})$\\
 
 SPad\'{e}-I &  $0.2806\pm 0.0086$  &  $0.6940^{+0.0066}_{-0.0077}$  &  $-$  &  $0.02\pm 0.13~(0.02\pm 0.25)$\\
 
 Pad\'{e}-II & $0.2762^{+0.0081}_{-0.0096}$  &  $0.6937\pm 0.0073$  &  $1.38^{+0.25}_{-0.21}~(1.38^{+0.48}_{-0.44})$  &  $-1.56\pm 0.18~(-1.56\pm 0.37)$\\
 
 $\Lambda$CDM & $0.2778\pm 0.0076$  &  $0.6946\pm 0.0071$  &  $-$  & $-$ \\
\hline 
\multicolumn{5}{c}{SN+BAO+BBN+$H(z)$+GW} \\
 \hline 

 Pad\'{e}-I & $0.261\pm 0.013$ & $0.6833\pm 0.0092$ & $-0.21\pm 0.54~(-0.21\pm 0.99)$ & $0.62\pm 0.56~(0.62\pm 1.08)$\\
 
SPad\'{e}-I & $0.258\pm 0.011$ & $0.6827\pm 0.0085$ & $-$ & $0.50^{+0.18}_{-0.15}~(0.50^{+0.38}_{-0.32})$   \\

 Pad\'{e}-II & $0.268\pm 0.014$ & $0.6856\pm 0.0095$ & $0.85^{+0.57}_{-0.67}~(0.85^{+1.07}_{-1.23})$ & $-1.1^{+0.71}_{-0.62}~(-1.1^{+1.4}_{-1.2})$\\
 
  $\Lambda$CDM & $0.2798\pm 0.0069$  &  $0.6893\pm 0.0074$  & $-$ & $-$\\

\hline 
\multicolumn{5}{c}{SN+BAO+BBN+$H(z)$+CMB+GW} \\
 \hline 

 Pad\'{e}-I & $0.2779^{+0.0079}_{-0.0091}$ & $0.6942\pm 0.0074$ & $-0.07^{+0.29}_{-0.26}~(-0.07^{+0.58}_{-0.54})$ & $0.12^{+0.32}_{-0.26}~(0.12^{+0.62}_{-0.54})$\\
 
SPad\'{e}-I & $0.2787\pm 0.0092$ & $0.6950\pm 0.0079$ & $-$ & $0.02^{+0.17}_{-0.21}~(0.02^{+0.33}_{-0.41})$\\

 Pad\'{e}-II & $0.2731\pm 0.0079$ & $0.6951^{+0.0065}_{-0.0080}$ & $0.60\pm 0.18~(0.60\pm 0.34)$ & $-0.77 \pm 0.21~(-0.77\pm 0.40)$\\
 
  $\Lambda$CDM & $0.2713\pm 0.0068$ & $0.6958\pm 0.0063$  & $-$ & $-$\\
\hline 
\end{tabular}\label{tab:flat}
\end{table*}

\begin{figure*}
\centering
\includegraphics[width=0.75\textwidth]{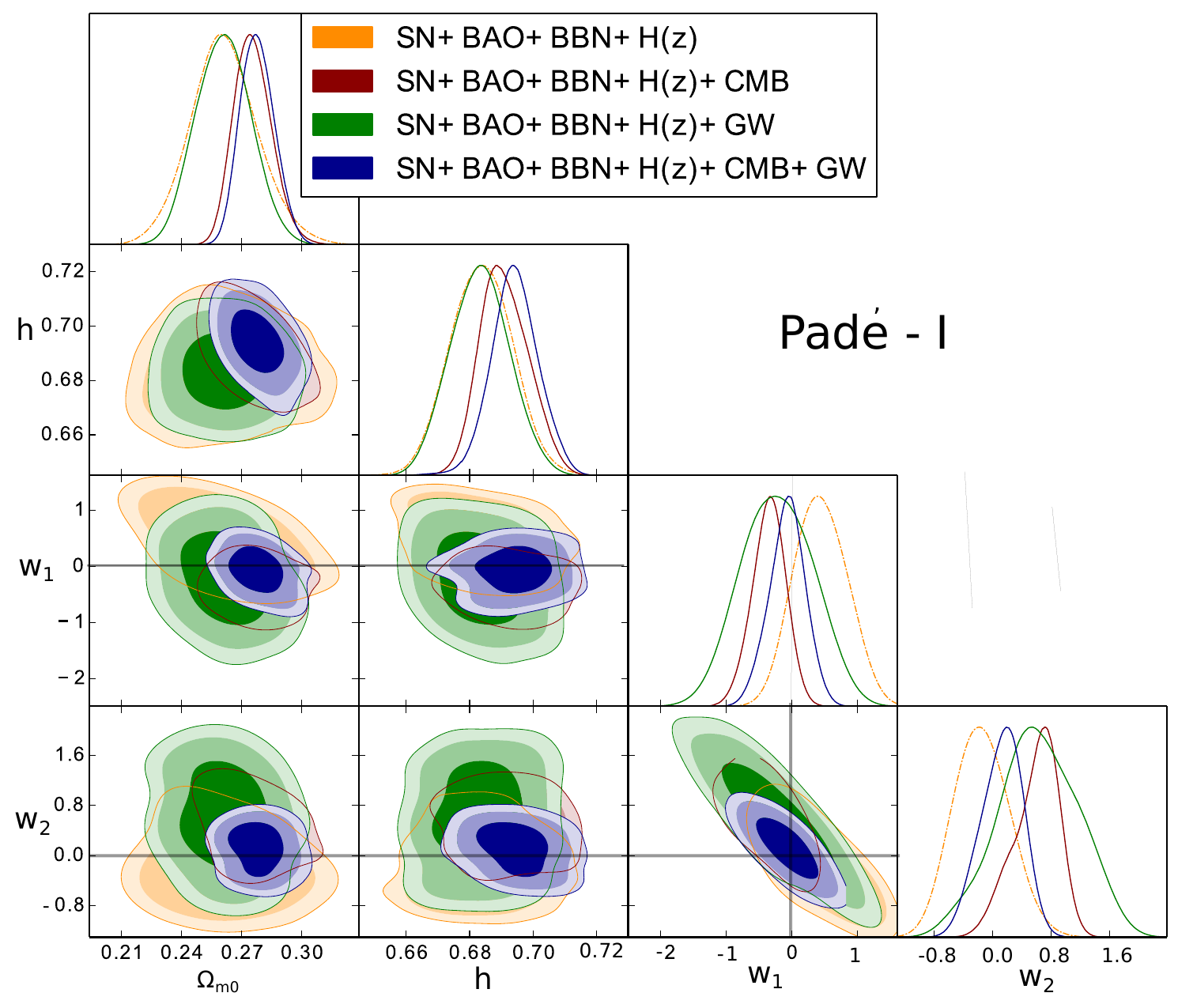}
\caption{One-dimensional marginalized posterior distributions and two-dimensional joint contours  at $1\sigma$ and $2\sigma$ considering some of the model parameters are shown for flat  Pad\'{e}-I using the combined datasets SN+BAO+BBN+$H(z)$, SN+BAO+BBN+$H(z)$+GW, SN+BAO+BBN+$H(z)$+CMB and SN+BAO+BBN+$H(z)$+CMB+GW. The solid lines through $w_1 =0$ and $w_2 = 0$ correspond to the $\Lambda$CDM cosmology, which is also considered as the fiducial model for generating the GW mock dataset. }
 \label{fig:p1k0}
\end{figure*}
\begin{figure*}
\centering
   \includegraphics[width=0.75\textwidth]{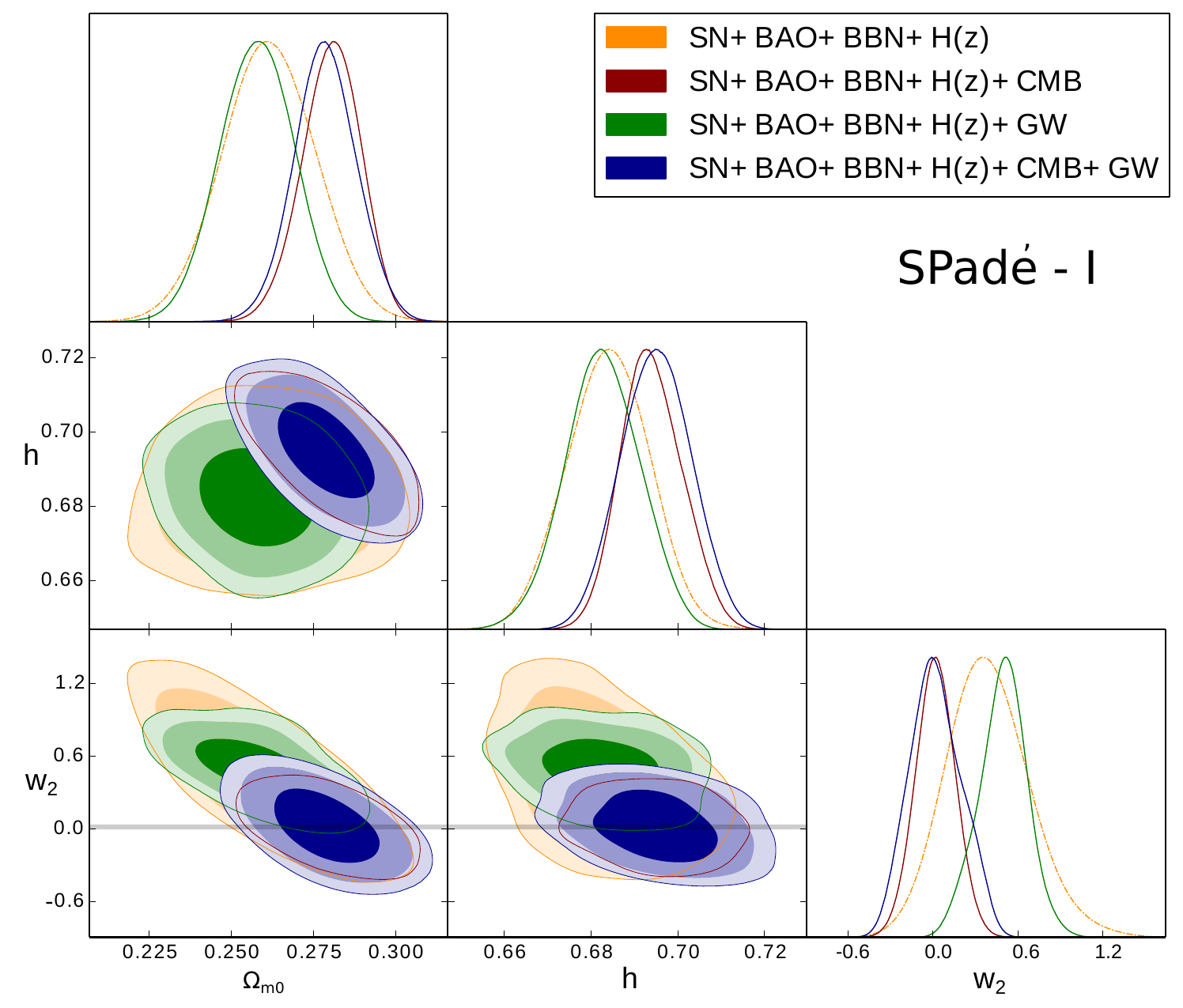}
 \caption{One-dimensional marginalized posterior distributions and two-dimensional joint contours  at $1\sigma$ and $2\sigma$ considering some of the model parameters are shown for flat SPad\'{e}-I using the combined datasets SN+BAO+BBN+$H(z)$, SN+BAO+BBN+$H(z)$+GW, SN+BAO+BBN+$H(z)$+CMB and SN+BAO+BBN+$H(z)$+CMB+GW. The solid line through  $w_2 = 0$ corresponds to the $\Lambda$CDM cosmology, which is also considered as the fiducial model for generating the GW mock dataset.    }
 \label{fig:spk0}
\end{figure*}
\begin{figure*}
\centering
\includegraphics[width=0.75\textwidth]{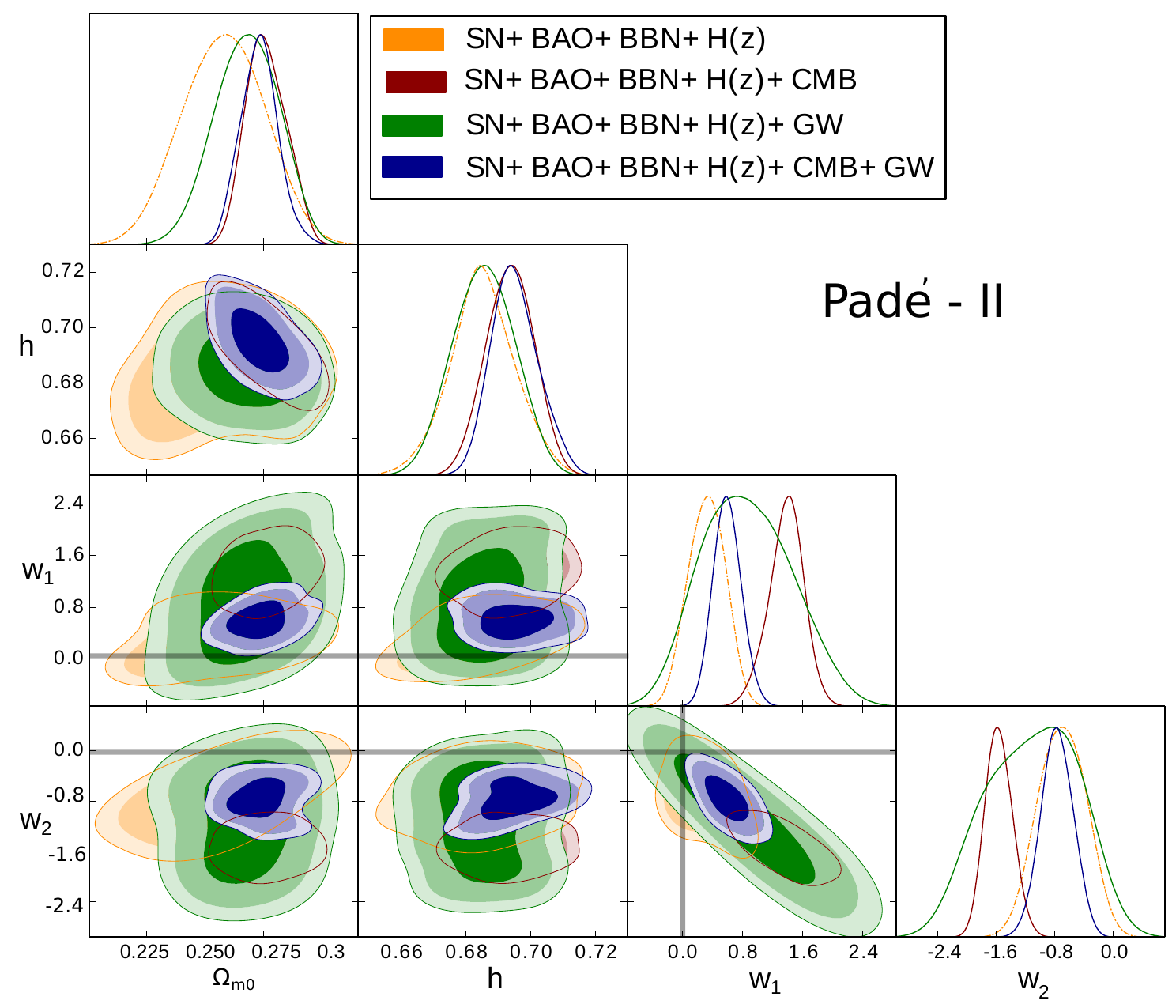}
\caption{One-dimensional marginalized posterior distributions and two-dimensional joint contours  at $1\sigma$ and $2\sigma$ considering some of the model parameters are shown for flat Pad\'{e}-II using the combined datasets SN+BAO+BBN+$H(z)$, SN+BAO+BBN+$H(z)$+GW, SN+BAO+BBN+$H(z)$+CMB and SN+BAO+BBN+$H(z)$+CMB+GW. The solid lines through $w_1 =0$ and $w_2 = 0$ correspond to the $\Lambda$CDM cosmology, which is also considered as the fiducial model for generating the GW mock dataset.    }
 \label{fig:p2k0}
\end{figure*}
\begin{table*}
\centering
\caption{Best-fit values of various free and derived parameters at $1\sigma$ ($2\sigma$) confidence level of the four cosmological scenarios in a nonflat universe considering the combined datasets. }
 \begin{tabular}{c  c  c c  c c }
 \hline 
 Model & $\Omega_{m0}$  & $\Omega_{k0}$  & $h$   & $w_1$ 
 & $w_2$\\
 \hline 

\multicolumn{6}{c}{SN+BAO+BBN+$H(z)$} \\
 \hline 

 Pad\'{e}-I & $ 0.2682\pm 0.0090$ & $0.008^{+0.041}_{-0.048}$ & $0.6868\pm 0.0082$  &  $-0.55\pm 0.41~(-0.55\pm 0.79)$  & $1.16\pm 0.60~(1.16\pm 1.18)$  \\
 
SPad\'{e}-I & $ 0.272\pm 0.014$ & $0.057\pm 0.039$  & $0.6875\pm 0.0099$ &  $-$ & $0.54^{+0.33}_{-0.28}$~($0.54^{+0.50}_{-0.53}$)\\

 Pad\'{e}-II & $ 0.270\pm 0.011$ & $-0.011\pm 0.018$ & $ 0.686^{+0.010}_{-0.0095}$ &  $0.56^{+0.07}_{-0.11}$~($0.56^{+0.18}_{-0.15}$) & $-0.30\pm 0.08$~($-0.30^{+0.18}_{-0.16}$) \\
 
  $\Lambda$CDM & $0.271\pm 0.015$ & $-0.017\pm 0.038$ & $0.6869\pm 0.0097$ &  $-$ & $-$ \\

\hline 
\multicolumn{6}{c}{SN+BAO+BBN+$H(z)$+CMB} \\
 \hline 
Pad\'{e}-I & $0.2760\pm 0.0080$ & $-0.001\pm 0.040$ & $0.6936\pm 0.0083$  &  $-0.64\pm 0.41 (-0.64\pm 0.79)$  & $0.78^{+0.67}_{-0.58}~(0.78 \pm 1.21)$  \\
 
 SPad\'{e}-I & $0.2776\pm 0.0087$ & $-0.009\pm 0.025$  & $0.6941\pm 0.0079$ &  $-$ & $0.34^{+0.10}_{-0.12}$~($0.34^{+0.24}_{-0.21}$) \\
 
 Pad\'{e}-II & $0.2778\pm 0.0083$ & $-0.026^{+0.022}_{-0.025}$ & $0.6934\pm 0.0078$ &  $0.48\pm 0.07~(0.48^{+0.14}_{-0.13}$)  & $-0.33 \pm 0.08$~($-0.33^{+0.15}_{-0.16}$) \\
 
 $\Lambda$CDM & $0.2768\pm 0.0081$ & $-0.0091\pm 0.0213$ & $0.6884\pm 0.0071$ & $-$ & $-$ \\
\hline 
\multicolumn{6}{c}{SN+BAO+BBN+$H(z)$+GW} \\
 \hline 
Pad\'{e}-I & $0.267^{+0.014}_{-0.012}$ & $0.029^{+0.021}_{-0.027}$ & $0.6855\pm 0.0097$ & $0.31\pm 0.35~(0.31\pm 0.67)$ & $0.09\pm 0.39~(0.09\pm 0.73)$ \\
 
 SPad\'{e}-I & $0.264\pm 0.014$ & $0.048\pm 0.049$ & $0.6861\pm 0.0099$ & $-$ & $0.65^{+0.43}_{-0.49}~(0.65^{+0.80}_{-0.93})$\\
 
 Pad\'{e}-II & $0.268\pm 0.016$ & $-0.022^{+0.038}_{-0.032}$ & $0.6855\pm 0.0092$ & $0.41^{+0.13}_{-0.10}~(0.41^{+0.19}_{-0.17})$ & $-0.43\pm 0.13~(-0.43\pm 0.22)$ \\
 
 $\Lambda$CDM & $0.2915^{+0.0091}_{-0.011}$ & $0.034^{+0.017}_{-0.022}$ & $0.6789\pm 0.0094$ &  $-$ & $-$ \\
\hline 
\multicolumn{6}{c}{SN+BAO+BBN+$H(z)$+CMB+GW} \\
 \hline 
Pad\'{e}-I & $0.2766^{+0.0083}_{-0.0094}$ & $0.102\pm 0.039$ & $0.6966\pm 0.0078$ & $-0.47^{+0.34}_{-0.38} (-0.47^{+0.64}_{-0.71})$ & $1.44^{+0.62}_{-0.36} (1.44^{+1.04}_{-0.75})$\\
 
 SPad\'{e}-I & $0.2755\pm 0.0088$ & $0.074\pm 0.055$ & $0.6954\pm 0.0080$ & $-$ & $0.66\pm 0.41 (0.66\pm 0.77)$
 \\
 
 Pad\'{e}-II & $0.279\pm 0.010$ & $-0.026\pm 0.020$ & $0.6949\pm 0.0082$ & $0.52^{+0.09}_{-0.08} (0.52^{+0.16}_{-0.15})$ & $-0.30\pm 0.06 (-0.30^{+0.11}_{-0.09})$ \\
 
 $\Lambda$CDM & $0.2841\pm 0.0073$ & $0.0159^{+0.0081}_{-0.027}$ & $0.6962\pm 0.0071$ & $-$ &  $-$ \\

\hline 
\end{tabular}\label{tab:non-flat}
\end{table*}

\begin{figure*}
\centering
   \includegraphics[width=0.75\textwidth]{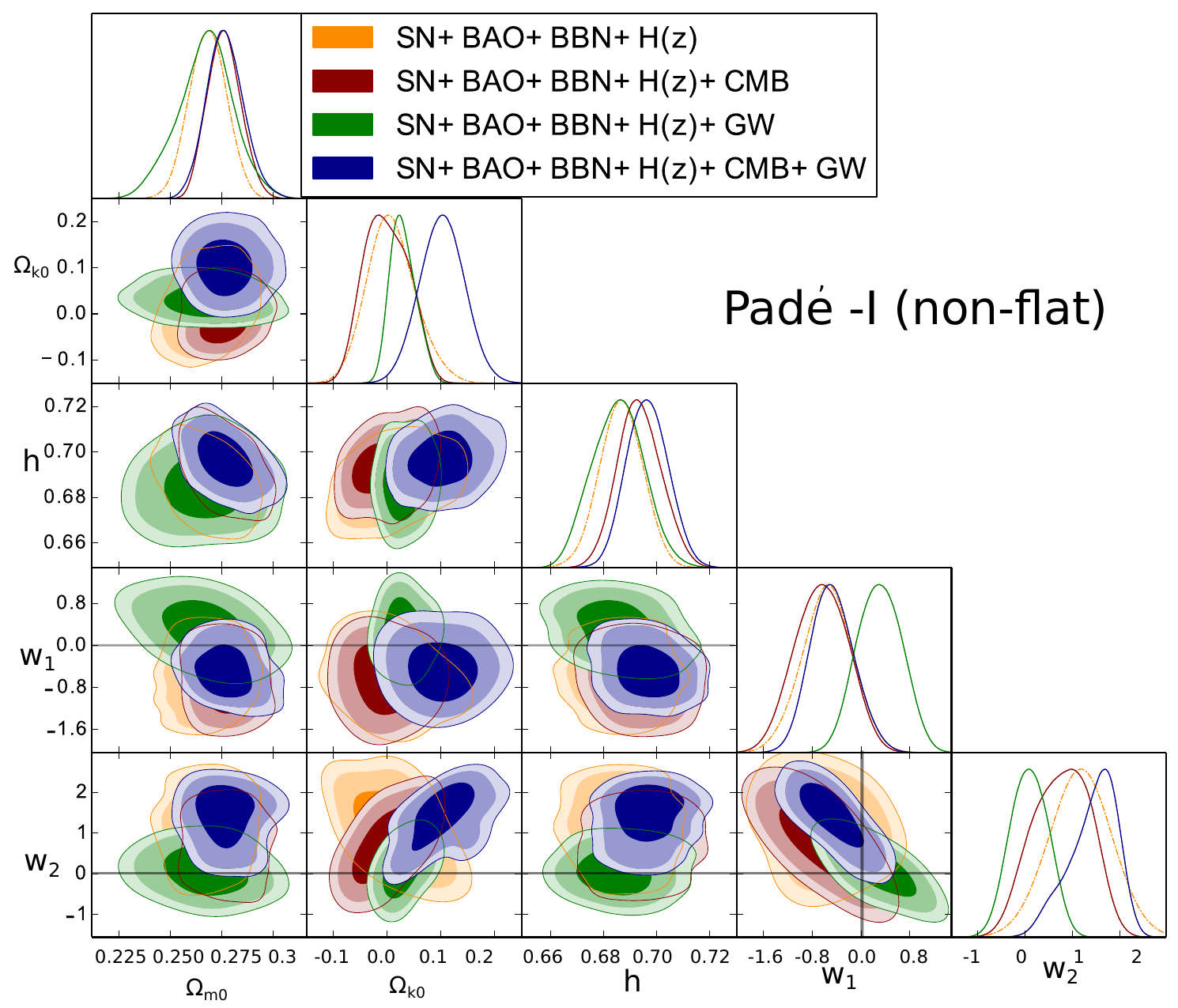}
   \caption{One-dimensional marginalized posterior distributions and two-dimensional joint contours  at $1\sigma$ and $2\sigma$ considering some of the model parameters are shown for non-flat Pad\'{e}-I using the combined datasets SN+BAO+BBN+$H(z)$, SN+BAO+BBN+$H(z)$+GW, SN+BAO+BBN+$H(z)$+CMB and  SN+BAO+BBN+$H(z)$+CMB+GW. The solid lines through $w_1 =0$ and $w_2 = 0$ correspond to the $\Lambda$CDM cosmology, which is also considered as the fiducial model for generating the GW mock data set.     }
 \label{fig:p1}
\end{figure*}
\begin{figure*}
\centering
   \includegraphics[width=0.75\textwidth]{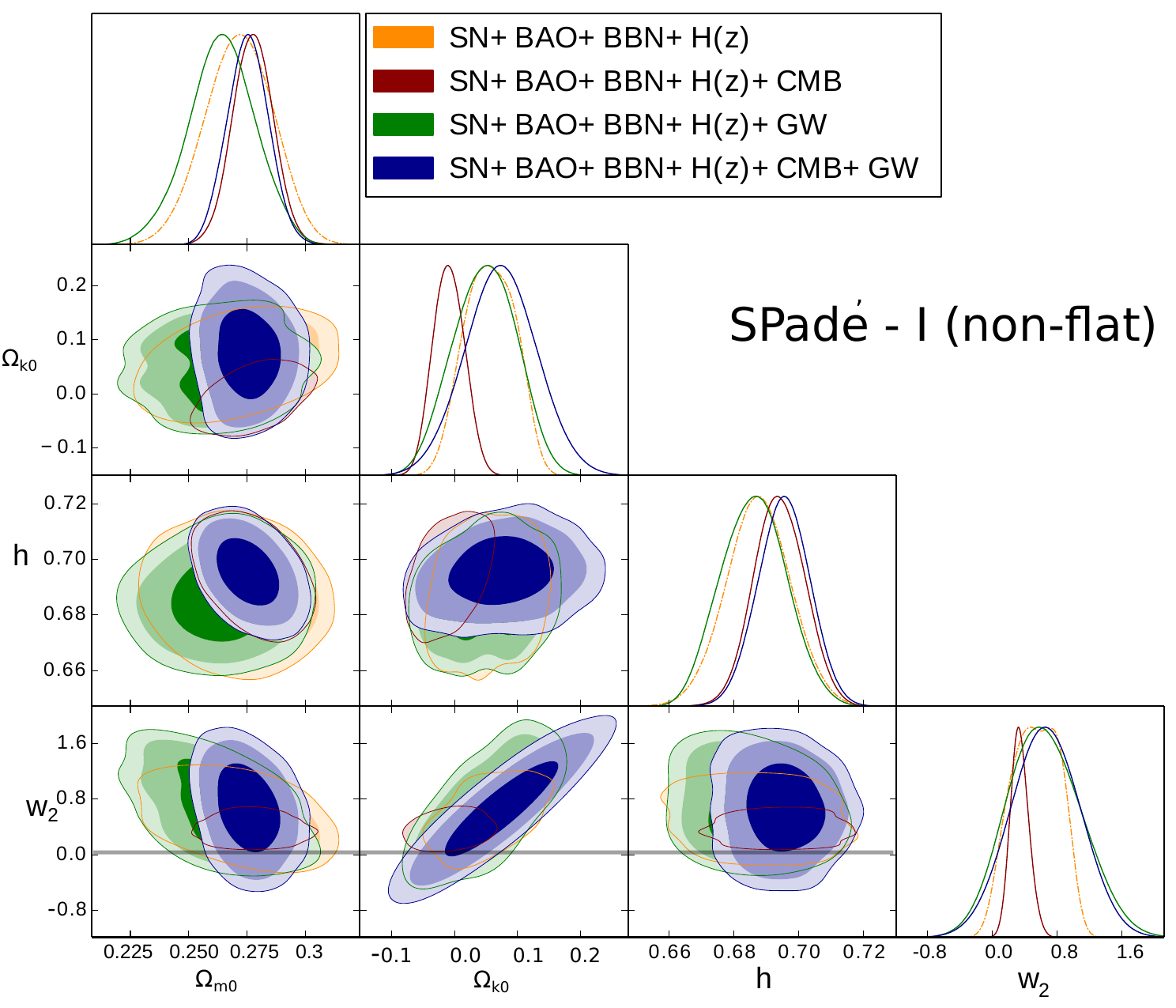}  
 \caption{One-dimensional marginalized posterior distributions and two-dimensional joint contours  at $1\sigma$ and $2\sigma$ considering some of the model parameters are shown for non-flat SPad\'{e}-I using  the combined datasets SN+BAO+BBN+$H(z)$, SN+BAO+BBN+$H(z)$+GW, SN+BAO+BBN+$H(z)$+CMB and SN+BAO+BBN+$H(z)$+CMB+GW. The solid line through $w_2 = 0$ corresponds to the $\Lambda$CDM cosmology, which is also considered as the fiducial model for generating the GW mock data set.        }
 \label{fig:sp}
\end{figure*}
\begin{figure*}
\centering
   \includegraphics[width=0.75\textwidth]{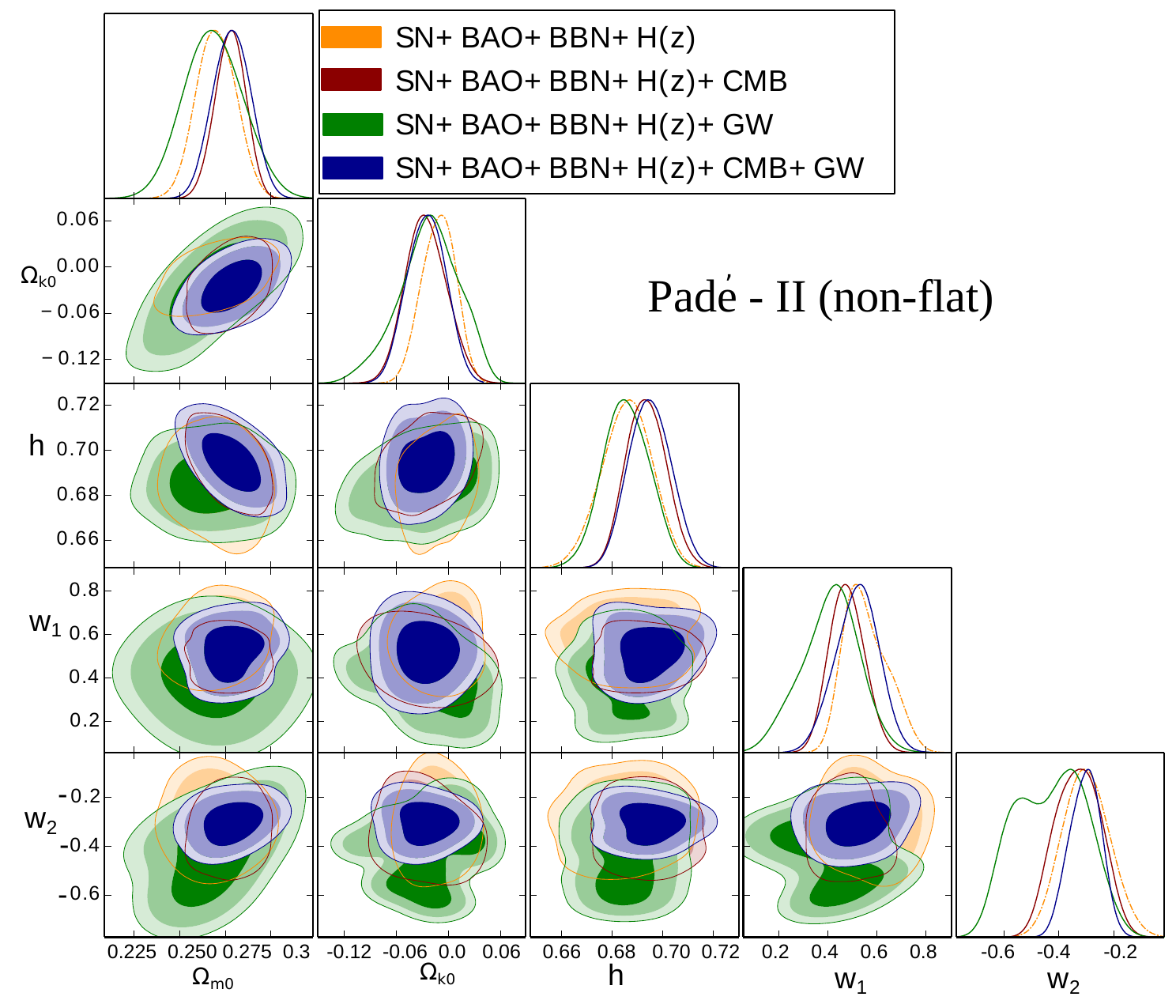}
\caption{One-dimensional marginalized posterior distributions and two-dimensional joint contours  at $1\sigma$ and $2\sigma$ considering some of the model parameters are shown for non-flat Pad\'{e}-II using the combined datasets SN+BAO+BBN+$H(z)$, SN+BAO+BBN+$H(z)$+GW, SN+BAO+BBN+$H(z)$+CMB and  SN+BAO+BBN+$H(z)$+CMB+GW. The solid lines through $w_1 =0$ and $w_2 = 0$ correspond to the $\Lambda$CDM cosmology, which is also considered as the fiducial model for generating the GW mock data set.    }
 \label{fig:p2}
\end{figure*}

\section{Results and their implications}
\label{sec-results}

In this section we summarize the observational constraints on the proposed Pad\'{e} parameterizations, namely, Pad\'{e}-I of eqn.  (\ref{pade1.2}), SPad\'{e}-I of eqn. (\ref{simplified-pade1.2}) and Pad\'{e}-II of eqn. (\ref{pade2.2}) for two separate cases: one for a vanishing curvature parameter and the other for a nonvanishing curvature parameter.  We have initially considered two combined datasets, namely, SN+BAO+BBN+$H(z)$ and SN+BAO+BBN+$H(z)$+CMB and then we have examined how the constraints are affected in the presence of the mock GWSS luminosity distance measurements. In Tables~\ref{tab:flat} and \ref{tab:non-flat} we present the constraints on the DE parameterizations for the spatially flat and nonflat universes, respectively. In Figs. \ref{fig:p1k0}, \ref{fig:spk0}, \ref{fig:p2k0}, \ref{fig:p1}, \ref{fig:sp}, \ref{fig:p2}, we summarize the graphical descriptions of the parametric scenarios. In particular, we show the one-dimensional posterior distributions of some key parameters and two-dimensional joint contours between some of the key parameters of these parametric scenarios. Finally, in Fig. \ref{fig:w} we present the evolution of $w_{\rm de}$ for all the Pad\'{e} parameterizations taking account of the combined probes of the real data sets.

\subsection{Constraints in a flat universe ($\Omega_k = 0$)} 

In Table~\ref{tab:flat} we present the observational constraints on the proposed Pad\'{e} parameterizations under the assumption of a spatially flat universe and considering various combined datasets, such as, (i) combined datasets taking into account of usual cosmological probes, namely, SN+BAO+BBN+$H(z)$ and SN+BAO+BBN+$H(z)$+CMB, and (ii) combined datasets in presence of the GW mock dataset, i.e., SN+BAO+BBN+$H(z)$+GW and SN+BAO+BBN+$H(z)$+CMB+GW.  
In Figs. \ref{fig:p1k0}, \ref{fig:spk0}, \ref{fig:p2k0} we show the one dimensional posterior distributions and two-dimensional joint contours for the flat Pad\'{e}-I, flat SPad\'{e}-I and flat Pad\'{e}-II, respectively, for the same combined datasets (with and without GW dataset). In the following, we present the observational constraints and their cosmological implications extracted out these Pad\'{e} parameterizations. Additionally, we note that in Table~\ref{tab:flat} we also present the constraints for the flat $\Lambda$CDM model acting as the reference model.  In what follows we present the constraints on each parameterization. 

We start with the constraints on Pad\'{e}-I obtained from the combined datasets of the usual cosmological probes, i.e., SN+BAO+BBN+$H(z)$ and SN+BAO+BBN+$H(z)$+CMB. From these datasets, we find that $w_1 \neq 0$ at more than $1\sigma$. In particular, the $1\sigma$ constraints on $w_1 $ are $w_1 = 0.44\pm 0.38$ (SN+BAO+BBN+$H(z)$) and $w_1 = -0.34^{+0.27}_{-0.24}$ (SN+BAO+BBN+$H(z)$+CMB). Regarding the remaining parameter, $w_2$, we notice that for  SN+BAO+BBN+$H(z)$, $w_2  = 0$ is allowed within $1\sigma$, but  the addition of CMB data to SN+BAO+BBN+$H(z)$, shows evidence of $w_2 \neq 0$ at more than $1\sigma$ ($w_2 = 0.60^{+0.37}_{-0.23}$ at $1\sigma$). Thus, considering both $w_1$ and $w_2$, we conclude that  mild evidence of a dynamical EoS of DE is supported by Pad\'{e}-I for both the combined datasets.  At this point, we recall the CPL parameterization ($w(a) = w_0 + w_a (1-a/a_0)$ \cite{Chevallier:2000qy, Linder:2002et}) since it also has two free parameters. However, it should be noted that CPL is not identical to Pad\'{e}-I in the sense that in Pad\'{e}-I, the present value of the DE EoS assumes $-1$ while in CPL, the current value of DE EoS, $w_0$, needs to be determined from the observational data. Nevertheless, one can examine the evidence for dynamical DE in different DE parameterizations.   As noted in the Planck 2018 paper~\cite{Planck:2018vyg}, Planck 2018 alone does not indicate any evidence for dynamical DE since in this case $w_a$ remains unconstrained. The inclusion of BAO from the Baryon Oscillation Spectroscopic Survey (BOSS) and SN data (Pantheon compilation) with Planck as performed by~\cite{Planck:2018vyg} also does not indicate evidence for dynamical DE. In particular, the 68\% CL constraints on $w_0$ and $w_a$ for Planck 2018+BAO are  
$w_0 = -0.957 \pm 0.080$ and $w_a = -0.29^{+0.32}
_{-0.26}$ (see Table~6 of \cite{Planck:2018vyg}). Thus, $w_a =0$ is allowed within the 68\% CL. This situation has changed a bit after the recent measurements of BAO from DESI collaboration~\cite{DESI:2024uvr}. According to DESI~\cite{DESI:2024uvr}, assuming CPL as the DE parameterization, evidence for  dynamical DE was reported at $\sim 2.6\sigma$ for the combined dataset DESI-BAO+CMB~\cite{DESI:2024mwx}, and this evidence increases when different SN samples are added to DESI-BAO+CMB~\cite{DESI:2024mwx}. 
When the GW dataset is added to both SN+BAO+BBN+$H(z)$ and SN+BAO+BBN+$H(z)$+CMB, we observe significant changes in the constraints. For example, for the combined dataset SN+BAO+BBN+$H(z)$+GW, the constraint on $w_1$ is now changed to $w_1 = -0.21 \pm 0.54$ at $1\sigma$ and the constraint on $w_2$ is now changed to $w_2 = 0.62 \pm 0.56$ at $1\sigma$. Thus,  significant shifts in the mean values of $w_1$ and $w_2$ are observed. However, $w_1 = 0$  is allowed within $1\sigma$ but $w_2 \neq 0$ is found at $1\sigma$. These jointly indicate mild evidence for dynamical DE at $1\sigma$ for the dataset SN+BAO+BBN+$H(z)$+GW. On the other hand, for the combined dataset SN+BAO+BBN+$H(z)$+CMB+GW, the constraints on both $w_1$ and $w_2$ are changed compared to  those for SN+BAO+BBN+$H(z)$+CMB, and hence the effects of the GW data set are visible. Despite these shifts, both $w_1$ and $w_2$ allow their null values within $1\sigma$. Hence, evidence for dynamical DE is not found in this case, but it should be noted that  uncertainties on the DE parameters are reduced after the inclusion of GW. Comparing the constraints from SN+BAO+BBN+$H(z)$+CMB and SN+BAO+BBN+$H(z)$+CMB+GW, one can readily observe that the uncertainties on both $w_1$ and $w_2$ are reduced, and this is because of GW.

We now discuss the constraints on the  SPad\'{e}-I parameterization.  For SN+BAO+BBN+$H(z)$, we find evidence of $w_2 \neq 0$ at more than $1\sigma$ ($w_2 = 0.39^{+0.27}_{-0.31} $at $1\sigma$), but this is consistent with $w_2 =0$ within $2\sigma$. After the addition of CMB to SN+BAO+BBN+$H(z)$ this evidence goes away ($w_2 = 0.02\pm 0.13$ at $1\sigma$) and $w_2$ becomes consistent with $w_2 =0$ within $1\sigma$. Thus, the evidence of dynamical EoS in this parameterization is not conclusive for the usual cosmological probes. However, when the GW dataset is added to both the above combined probes, we notice that for SN+BAO+BBN+$H(z)$+GW, $w_2$ is found to be nonzero at more than $2\sigma$ ($w_2 = 0.50^{+0.38}_{-0.32}$ at $2\sigma$), but for SN+BAO+BBN+$H(z)$+CMB+GW, we do not have any evidence of $w_2 \neq 0$. In fact,  for SN+BAO+BBN+$H(z)$+CMB+GW, $w_2 =0$ is consistent within $1\sigma$. Thus, we see that when CMB is taken into account, we do not find any evidence of dynamical DE within this parametrized framework.

Finally we focus on the last parameterization, i.e.  Pad\'{e}-II.  For the combined dataset SN+BAO+BBN+$H(z)$, we find that $w_1$ is non-null at $1.4\sigma$ and $w_2$ is non-null at $1.6\sigma$, hence, a mild evidence of dynamical DE is suggested in this case.  When the CMB data are added to SN+BAO+BBN+$H(z)$, we observe that both $w_1 \neq 0$ at $4.8\sigma$ and $w_2 \neq 0$ at $6.4\sigma$ which clearly indicates strong evidence of a dynamical EoS in DE (see Fig. \ref{fig:p2k0}).  This is a very interesting result in the context of dynamical DE because according to recent DESI-BAO measurements, assuming CPL as the DE parameterization, evidence of dynamical DE has been found at $\sim 2.6\sigma$ for the DESI-BAO+CMB~\cite{DESI:2024mwx}, and this evidence can increase up to $3.9\sigma$ depending on the underlying SN dataset~\cite{DESI:2024mwx}. Although CPL and  Pad\'{e}-II  are structurally different, both of them are heading toward the dynamical evidence of DE.  
Additionally, apart from CPL, evidence for dynamical DE has also been found in a class of well known DE parameterizations~\cite{Giare:2024gpk}. The results from CPL and other DE parameterizations strengthen  Pad\'{e}-II as an emerging DE candidate.  
In the presence of the GW dataset, the evidence for dynamical DE is more pronounced when CMB data are present in the combined dataset. That means, as we notice, for SN+BAO+BBN+$H(z)$+GW, the evidence for dynamical DE is found at slightly more than $1\sigma$ (hence, mild evidence is noted),  however,  for SN+BAO+BBN+$H(z)$+CMB+GW, this evidence is again restored at many standard deviations ($w_1 \neq 0$ at $3.1\sigma$ and $w_2 \neq 0$ at $4.7\sigma$).  In addition to that, one can notice that the uncertainties on both $w_1$ and $w_2$ are reduced after the addition of GW to SN+BAO+BBN+$H(z)$+CMB. This is an interesting observation in this model since the evidence of dynamical DE is restored and the uncertainties on the DE parameters are also reduced.

\subsection{Constraints in a nonflat universe: treating $\Omega_k$ as a free-to-vary parameter}

In comparison with the results presented in the previous section, here one of the new ingredients is the curvature of the universe. 
In Table~\ref{tab:non-flat} we display the observational constraints on the proposed  Pad\'{e} parameterizations and the $\Lambda$CDM paradigm in the nonflat background of the universe. In a similar fashion, we have tested all these scenarios using two combined datasets taking account of the usual cosmological probes, namely, SN+BAO+BBN+$H(z)$ and SN+BAO+BBN+$H(z)$+CMB, and finally we added the mock GW dataset to both the above datasets and constrained the resulting DE scenarios.  In Figs. \ref{fig:p1}, \ref{fig:sp}, \ref{fig:p2}, we  show the one-dimensional posterior distributions and two-dimensional joint contours for the nonflat Pad\'{e}-I, nonflat SPad\'{e}-I and nonflat Pad\'{e}-II, respectively, for the same datasets. 
In what follows we present the main results obtained in the scenarios. 

We begin with the Pad\'{e}-I parameterization. Considering the constraints from SN+BAO+BBN+$H(z)$ and SN+BAO+BBN+$H(z)$+CMB, first of all we notice that  $w_1$ and $w_2$ are nonzero at more than $1\sigma$ for both the datasets, while they take null values within $2\sigma$. Thus,  a dynamical nature in $w(a)$ describing Pad\'{e}-I is mildly allowed. On the other hand, regarding the curvature of the universe, 
we do not find any strong evidence of a non zero value of  $\Omega_{k, 0}$  and $\Omega_{k0} =0$ is consistent with its zero value within $1\sigma$.  
When the GW dataset is added to both SN+BAO+BBN+$H(z)$, SN+BAO+BBN+$H(z)$+CMB, regarding the curvature of the universe,  mild evidence of its nonzero value cannot be avoided for SN+BAO+BBN+$H(z)$+GW since $\Omega_{k0} = 0.029^{+0.021}_{-0.027}$ at $1\sigma$ in this case, while SN+BAO+BBN+$H(z)$+CMB+GW does not indicate such evidence. On the other hand, quite notably, the constraints on $w_1$ and $w_2$ significantly change in the presence of the GW dataset. In particular, the change of the constraints is clearly visible for SN+BAO+BBN+$H(z)$+GW compared to SN+BAO+BBN+$H(z)$+CMB+GW.  As displayed in Table~\ref{tab:non-flat}, for SN+BAO+BBN+$H(z)$+GW, we do not find any evidence of a dynamical nature in DE, since $w_1$ and $w_2$ are consistent to their null values within $1\sigma$. However, things change for SN+BAO+BBN+$H(z)$+CMB+GW where we notice that $w_1 \neq 0$ at more than $1\sigma$ and $w_2 \neq 0 $ at more than $2\sigma$. Thus,  evidence of a dynamical nature in the DE EoS and consequently a deviation from $w(a) =-1$ is pronounced for SN+BAO+BBN+$H(z)$+CMB+GW.

For the SPad\'{e}-I parameterization,  according to the combined datasets SN+BAO+BBN+$H(z)$ and SN+BAO+BBN+$H(z)$+CMB, $w_2 \neq 0$ at more than $2\sigma$, hence evidence of a dynamical DE together with a deviation from $w =-1$ is inferred. Concerning the curvature parameter,  an evidence of $\Omega_{k0} \neq 0$ at more than $1\sigma$ ($\Omega_{k, 0} = 0.54^{+0.33}_{-0.28}$ at $1\sigma$) for SN+BAO+BBN+$H(z)$ is found, but after the inclusion of CMB to SN+BAO+BBN+$H(z)$, this evidence goes away and $\Omega_{k0}$ is consistent with a spatially flat universe ($\Omega_{k0} = -0.009\pm 0.025$ at $1\sigma$ for SN+BAO+BBN+$H(z)$+CMB).  
The addition of the GW dataset to SN+BAO+BBN+$H(z)$ and SN+BAO+BBN+$H(z)$+CMB mildly affects DE parameters. For example, $w_1$ and $w_2$ are currently now non null at more than $1\sigma$, but notice that these two parameters  were previously found to be non zero at more than $2\sigma$ without adding the GW dataset. Thus, after the inclusion of GW, the evidence of dynamical DE is slightly weakened. Regarding the curvature of the universe, we find mild evidence of $\Omega_{k0} \neq 0$ for 
SN+BAO+BBN+$H(z)$+GW. Interestingly, we notice that $\Omega_{k0}$ which was consistent with zero  for SN+BAO+BBN+$H(z)$+CMB, is found to be nonzero for SN+BAO+BBN+$H(z)$+CMB+GW, and specifically an indication of a open universe ($\Omega_{k0} = 0.074  \pm 0.055$ at $1\sigma$ for SN+BAO+BBN+$H(z)$+CMB+GW) is found in this case.

Finally we consider the last parameterization of this series, namely, Pad\'{e}-II. In this case,  for both SN+BAO+BBN+$H(z)$ and SN+BAO+BBN+$H(z)$+CMB, we find that both the DE parameters, namely, $w_1$ and $w_2$ are nonzero at several standard deviations. In particular, for SN+BAO+BBN+$H(z)$, we find that $w_1$ takes the null value  at $6.2 \sigma$ and $w_2$ takes the null value at $3.3\sigma$. This clearly indicates strong evidence of a dynamical DE, and $\Lambda$CDM is excluded at $6.2 \sigma$. For SN+BAO+BBN+$H(z)$+CMB, the evidence for dynamical DE is more pronounced and in this case the null values of $w_1$ and $w_2$ are located at $6.4\sigma$ and $4.8\sigma$, respectively. In other words, it can be stated that this scenario rejects  $w(a) =-1$ with a very high confidence level. This is in agreement with the recent results by the DESI collaboration assuming CPL ~\cite{DESI:2024mwx} and other DE parameterizations~\cite{Giare:2024gpk, Giare:2024oil}. The only difference is that,  within Pad\'{e}-II, we find a very strong preference for dynamical DE over CPL and other DE parameterizations explored so far~\cite{DESI:2024mwx,Giare:2024gpk, Giare:2024oil}.  
Additionally, we find a mild evidence of a closed universe~($\Omega_{k0} = -0.026^{+0.022}_{-0.025}$ at $1\sigma$) for the combined dataset SN+BAO+BBN+$H(z)$+CMB.  

In the presence of the GW dataset, our conclusions do not change. We again notice a strong indication of a dynamical DE, henceforth a deviation from $w(a) =-1$ at several standard deviations, and evidence of a closed universe for SN+BAO+BBN+$H(z)$+CMB+GW leading to $\Omega_{k0} = -0.026 \pm 0.020$ at $1\sigma$, is found. In particular, for SN+BAO+BBN+$H(z)$+GW, the null values of $w_1$ and $w_2$ are attained at $2.7\sigma$ and $3.8 \sigma$, respectively. On the other hand, in the presence of CMB, which means for the combined dataset SN+BAO+BBN+$H(z)$+CMB+GW, $w_1$ and $w_2$ attain null values at $5.1\sigma$ and $4.2\sigma$, respectively. These clearly indicate that we have very significant support for dynamical evolution of DE in the Pad\'{e}-II scenario assuming all of our data samples.

\begin{figure*}
\includegraphics[width=0.49\textwidth]{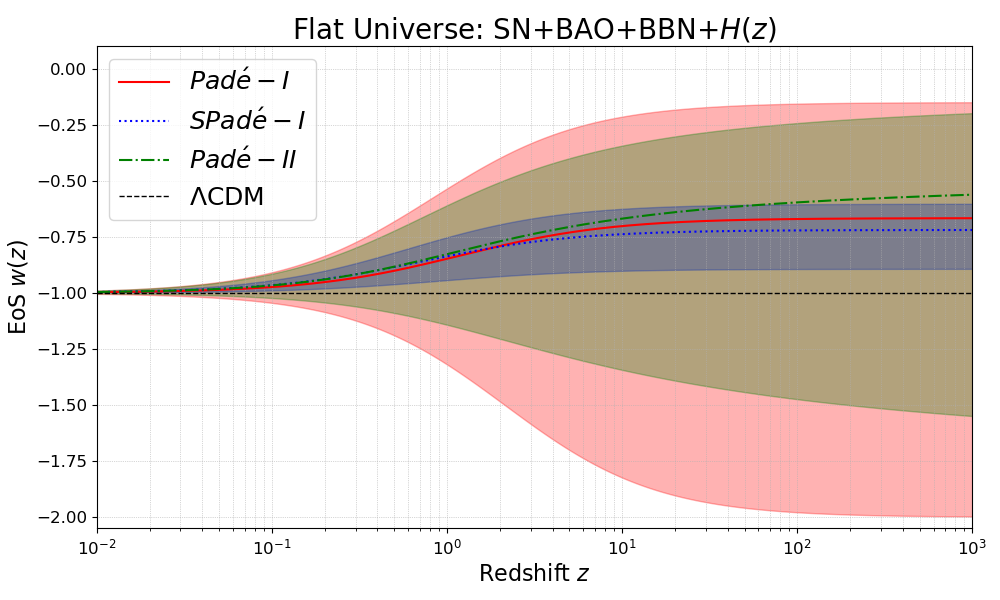}
\includegraphics[width=0.49\textwidth]{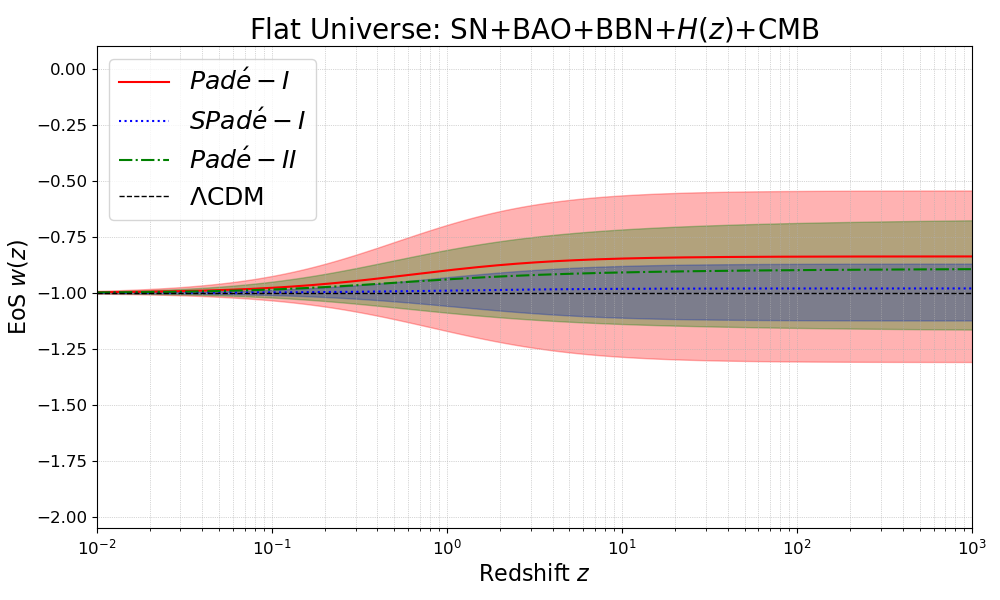}
\includegraphics[width=0.49\textwidth]{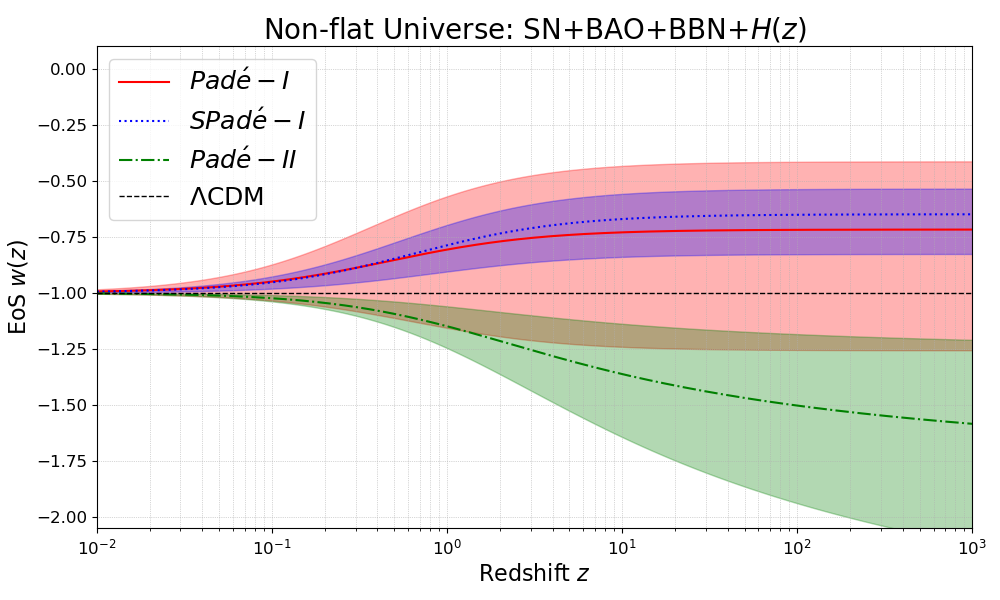}
\includegraphics[width=0.49\textwidth]{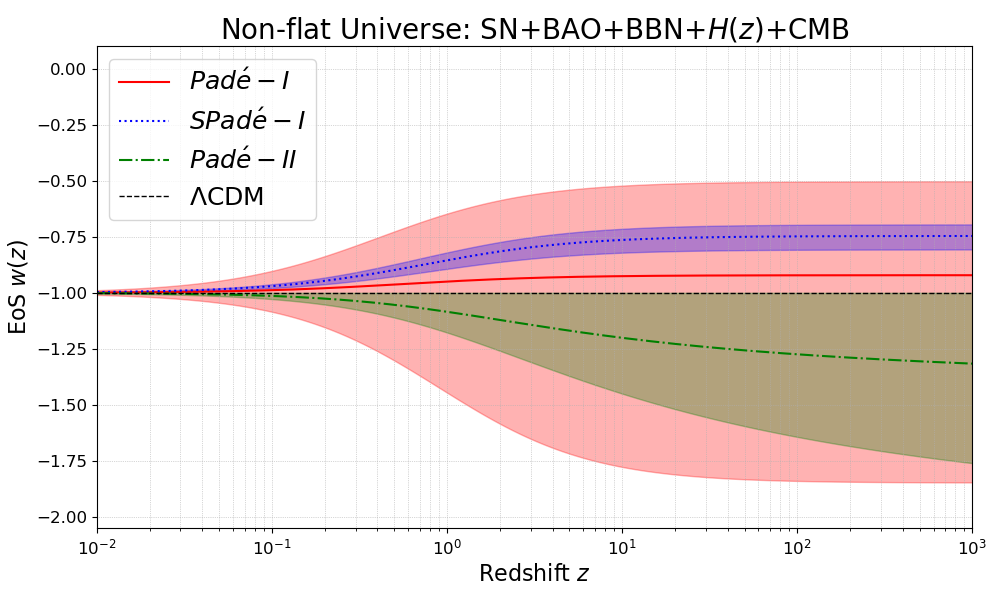}
\caption{The redshift evolution of $w (z)$ with its 68\% CL region for different DE parameterizations (Pad\'{e}-I, SPad\'{e}-I, Pad\'{e}-II). In the upper panels we show the evolution of $w(z)$ assuming the flat universe while in the lower panels we show the evolution of $w(z)$ considering the non-flat universe. 
The left panels corresponds to SN+BAO+BBN+$H(z)$ and the right  panels corresponds to SN+BAO+BBN+$H(z)$+CMB.  }
 \label{fig:w}
\end{figure*}

We now close this section with Fig. \ref{fig:w} which shows the evolution of $w(z)$ with its 68\% CL region for all the Pad\'{e} parameterizations, taking into account the constraints from two combined datasets, namely, SN+BAO+BBN+$H(z)$ and SN+BAO+BBN+$H(z)$+CMB in the spatially flat universe (see Table~\ref{tab:flat}) and nonflat universe (see Table~\ref{tab:non-flat}). 
Focusing on the 
flat case (the upper panel of Fig. \ref{fig:w}), our first observation is that the addition of CMB to SN+BAO+BBN+$H(z)$ significantly affects the parameter space. For SN+BAO+BBN+$H(z)$, the mean curves of $w(z)$ in the high-redshift regime irrespective of the parameterizations strictly lie in the quintessence regime, but $w(z) < -1$ is also allowed within the 68\% CL  (except for SPad\'{e}-I). For SN+BAO+BBN+$H(z)$+CMB (flat case), we see that within the 68\% CL, $w(z)$ allows both the quintessence and phantom regimes, although the mean curves of $w(z)$ for all the parameterizations lie in the quintessence regime. One can further notice that the mean curve of $w(z)$ for SPad\'{e}-I almost mimics $w(z) =-1$.  
We now turn to the lower panel of Fig. \ref{fig:w} which corresponds to the nonflat case. Comparing again the two graphs in the lower panel, our first observation is that the addition of CMB to SN+BAO+BBN+$H(z)$ 
similarly affects the parameter space, but not significantly, as observed in the flat case (see the upper panel of Fig. \ref{fig:w}). In this case, we see that for SPad\'{e}-I, the high-redshift evolution of $w(z)$  remains quintessential at more than the  68\% CL irrespective of the combined datasets, while for Pad\'{e}-I, although the mean curve of $w(z)$ remains in the quintessential regime for both the combined datasets, but within the 68\% CL region, a phantom nature of $w(z)$ is not ruled out. On the other hand, Pad\'{e}-II exhibits some different results: for SN+BAO+BBN+$H(z)$, evolution of $w(z)$ in the high-redshift regime remains in the phantom regime at more than the 68\% CL, and with decreasing $z$, $w(z)$ approaches the cosmological constant boundary;  for SN+BAO+BBN+$H(z)$+CMB, the mean curve of $w(z)$  remains in the phantom regime, but it marginally allows $w(z) =-1$.

\begin{table*}
\centering
\caption{Statistical results of Bayesian evidence for different cosmological models studied in this work. Here we have $\Delta \ln \epsilon= \ln \epsilon_{\Lambda CDM}-\ln \epsilon_{Model}$ for each of the scenarios.}
 \begin{tabular}{c  c  c c c}
 \hline 
 & \multicolumn{2}{c}{SN+BAO+BBN+$H(z)$} & \multicolumn{2}{c}{SN+BAO+BBN+$H(z)$+CMB} \\
 \hline 
 Model & flat universe  & non-flat universe  & flat universe  & non-flat universe\\
 \hline
 Pad\'{e}-I &  $0.8$  &  $0.4$  &  $0.9$  & $0.6$  \\
 
SPad\'{e}-I &  $0.1$   & $1.1$  & $-0.1$  & $0.3$ \\

 Pad\'{e}-II &  $0.8$  &  $1.0$  &  $0.7$  & $0.8$  \\
 
\hline 
\end{tabular}\label{tab:BE}
\end{table*}

\section{Model Comparison}
\label{sec-model-comparison}

In this section we consider two model comparison methods, namely Bayesian evidence analysis and the information criteria  which are frequently used to assess the fitness of cosmological models with the data. In the first half of this section we present the Bayesian evidence analysis for the proposed Pad\'{e} parameterizations and the last part describes the inference from several information criteria. For both methodologies, one needs to prescribe a reference model with respect to which the ability of the models is tested; in this article,
$\Lambda$CDM has been treated as the reference model. Since we have studied our DE parameterizations in both flat and non-flat universes, we have taken the flat $\Lambda$CDM model as the reference model for the flat Pad\'{e} parameterizations and for the nonflat $\Lambda$CDM model as the reference model for the nonflat DE models.  We now discuss the Bayesian evidence analysis in the following. 
Assuming $\Theta$ being the free parameters of model $M$, and $D$ as the observational data samples, the Bayesian evidence $\epsilon$, is given by \cite{Rezaei:2021qpq}:
\begin{eqnarray}
\epsilon = p(D \mid M)=\int p(\Theta \mid M)p(D \mid \Theta,M)d\Theta\;.
\end{eqnarray}
The evidence is a crucial quantity for model selection in the  Bayesian framework. Among different models, a model with higher evidence is favored over another \cite{Rezaei:2020mrj}. Here, we use the Jeffrey's scale \cite{jeffreys1983theory} to measure the significant difference between different models. Considering two models $M_1$ and $M_2$, the Jeffreys scale in respect of $\Delta \ln \epsilon = \ln \epsilon_{ M_1} -\ln \epsilon_{ M_2}$ is as follows \cite{Nesseris:2012cq}:

\begin{itemize}
\item for $\Delta \ln \epsilon < 1.1$ we have weak evidence against $M_2$.
\item for $1.1<\Delta \ln \epsilon < 3$ we have definite evidence against $M_2$.
\item for $\Delta \ln \epsilon > 3$ we have strong evidence against $M_2$ 
\end{itemize}
The numerical results of this part are reported in Table~\ref{tab:BE}. In this table we have reported $\Delta \ln \epsilon$ for each of the Pad\'{e} parameterization computed with respect to the flat (or nonflat) $\Lambda$CDM, which is defined as $\Delta \ln \epsilon =$ $\ln \epsilon$ {$\Lambda$CDM} $-$ $\ln \epsilon$ {\rm model}.\footnote{The negative sign of $\Delta \ln \epsilon$  indicates that the model is preferred over the $\Lambda$CDM while the positive sign denotes the opposite. }  We see that flat SPad\'{e}-I is found to be the best model among all three Pad\'{e} parameterizations. This becomes prominent when CMB data are added to the combined dataset. In fact, what is more interesting to note here is that SPad\'{e}-I is favored over $\Lambda$CDM when CMB is present in the combined dataset ($\ln \epsilon =-0.1$), but this preference is not very strong.   
Overall, we see that $\Lambda$CDM remains the best model among all three parameterizations, while we have weak evidence against Pad\'{e} parameterizations. 
As the main result of this part, we observed that there is no significant difference between our Pad\'{e} parameterizations and $\Lambda$CDM cosmology from the  point of view of Bayesian evidence. Further, since in all cases we have $\Delta \ln \epsilon \leq  1.1$,  one can safely conclude that all evidence is weak, and it is very hard to prefer one parameterization over other.

\begin{table*}
\centering
\caption{Summary of $\chi^2_{\rm min}$ values obtained for different models $\Delta \chi^2_{\rm min}$ $=$ $\chi^2_{\rm min}$ (model) $-$  $\chi^2_{\rm min}$ ($\Lambda$CDM) 
and different information criteria including AIC, BIC and DIC for Pad\'{e} parameterizations with respect to the reference model $\Lambda$CDM (flat and nonflat). We note that in all cases, $\Delta {\rm XIC}=$ XIC (model) $-$ XIC ($\Lambda$CDM).  The negative sign of $\Delta$XIC  indicates that the model is preferred over the $\Lambda$CDM while the positive sign denotes the opposite.}  
 \begin{tabular}{c c c  c c  c c }
 \hline 
 Curvature & Model & $\chi^2_{\rm min}$  & $\Delta \chi^2_{\rm min}$ & $\Delta {\rm AIC}$  & $\Delta {\rm BIC}$   & $\Delta {\rm DIC}$\\
 \hline 

\multicolumn{7}{c}{SN+BAO+BBN+$H(z)$} \\
 \hline 

 		& Pad\'{e}-I & 1058.5  & $-3.0$ & 1.0 & 3.1 & 1.2\\
 
$\Omega_{k0}=0$ & SPad\'{e}-I & 1057.9  & $-3.6$ & $-1.6$ & $-0.6$ & $-0.2$ \\

 		& Pad\'{e}-II & 1061.2 & $-0.3$ & 3.7 & 5.8 & 0.0 \\
 
 		& $\Lambda$CDM & 1061.5 & $-$ & $-$ & $-$ & $-$ \\
\hline 

 		& Pad\'{e}-I & 1057.6  & $-1.7$ & 2.3 & 4.4 & 0.6 \\
 
$\Omega_{k0}\neq 0$ & SPad\'{e}-I & 1057.4  & $-1.9$ & 0.1 & 1.1 & 0.0 \\

 		& Pad\'{e}-II & 1061.0 & 1.7 & 5.7 & 7.8 & 0.8 \\
 
 		& $\Lambda$CDM & 1059.3 & $-$ & $-$ & $-$ & $-$\\

\hline 
\multicolumn{7}{c}{SN+BAO+BBN+$H(z)$+CMB} \\
 \hline 

 		& Pad\'{e}-I & 1061.2 & $-0.9$ & 3.1 & 5.2 & 0.9\\
 
$\Omega_{k0}=0$ & SPad\'{e}-I & 1058.5 & $-3.6$ & $-1.6$ & $-0.6$ & 0.1 \\

 		& Pad\'{e}-II & 1061.4 & $-0.7$ & 3.3 & 5.4 & 0.0\\
 
 		& $\Lambda$CDM & 1062.1 & $-$ & $-$ & $-$ & $-$ \\
\hline 

 		& Pad\'{e}-I & 1061.0 & $-0.4$ & 3.6 & 5.7 & 0.5 \\
 
$\Omega_{k0} \neq 0$ & SPad\'{e}-I & 1058.0 & $-3.4$ & $-1.4$ & $-0.4$ & $-0.8 $\\

 		& Pad\'{e}-II & 1061.5 & 0.1 & 4.1 & 6.2 & 1.1  \\
 
 		& $\Lambda$CDM & 1061.4 & $-$ & $-$ & $-$ & $-$ \\

\hline 
\end{tabular}\label{tab:IC}
\end{table*}

We now present various information criteria that have been used to test the current Pad\'{e} parameterizations. We have used three  information criteria that are frequently used in cosmology, namely, the Akaike Information Criterion (AIC) \citep{Akaike:1974vps}, the Bayesian Information Criterion (BIC) \citep{Schwarz:1978tpv} and Deviance Information Criterion (DIC) \citep{Spiegelhalter:2002yvw}. The AIC and BIC have the following forms: 
\begin{eqnarray}
&&{\rm AIC} = \chi^2_{\rm min}+2M,\\ 
&& {\rm BIC} = \chi^2_{\rm min}+M \ln N\;,
\end{eqnarray} 
where $M$ is the total number of parameters of the model considered in the observational tests and $N$ denotes the total number of data points in the observational analysis. The DIC, on the other hand,  assumes a different expression as follows
\citep{Liddle:2007fy}
\begin{eqnarray}
{\rm DIC} = D(\bar{{\bf p}})+2Y_{\rm B},
\end{eqnarray}
where $Y_{\rm B}=\overline{{D({\bf p})}}-D(\bar{{\bf p}})$ 
is the Bayesian complexity, ${\bf p}$ is the vector of parameters and overlines imply the standard  mean value. We note that $D(\bar{{\bf p}})$ represents the minimum value of $\chi^2$, while $\overline{{D({\bf p})}}$ implies the mean value of all $\chi^2$ values that we obtained in our markov chain more carlo chains. One can notice that DIC is different than AIC and BIC because it includes both the Bayesian statistics and the concept of information theory~\citep{spiegelhalter2002bayesian}.   
Here $D({\bf p})$ represents the Bayesian deviation, which takes the form $D({\bf p})=\chi^2_{\rm t}({\bf p})$ in the case of an exponential class of distributions (see ~\cite{Trashorras:2016azl,Rezaei:2021qpq} for more details). Now, as already mentioned, one needs to consider a reference model in order to examine the  fitness of the underlying models with the data, and $\Lambda$CDM has been considered as the reference model. We compute the
differences $\Delta$XIC $=$ XIC (for model) $-$ XIC (for $\Lambda$CDM), where X$=$ A, B, D. Note that when we consider the Pad\'{e} parameterizations in the flat universe, we choose flat $\Lambda$CDM as the reference model, and when we consider the Pad\'{e} parameterizations in the nonflat universe, then we change our reference model from flat to nonflat $\Lambda$CDM.  Using the values of $\Delta {\rm XIC}$, we can evaluate the level of support  for (or the strength of evidence against) a given DE model. Concerning AIC and DIC, $\Delta {\rm XIC}<2$ means we have ``significant support'' for the given model; $4<\Delta {\rm XIC} <7$ indicates ``considerably less support'' for the given model, and $10<\Delta {\rm XIC}$ implies ``essentially no support'' for the given model. In the same manner, for BIC, $\Delta {\rm BIC}<2$ shows ``no evidence'' against the model; $2<\Delta {\rm BIC}<6$ indicates ``mild to positive evidence'' against the model, and  $6<\Delta {\rm BIC}<10$ indicates ``strong evidence'' against the model, and $\Delta {\rm BIC}> 10$  means ``very strong evidence'' against the model (see Tables~II and III of \citep{Rezaei:2019xwo}). We note that all of these intervals are merely general rules of thumb commonly used in the literature in order to assess the viability of a model with respect to the reference one. The negative sign of $\Delta$XIC  indicates that the model is preferred over the $\Lambda$CDM while the positive sign denotes the opposite.  
In Table~\ref{tab:IC} we summarize the results of the information criteria for all the Pad\'{e} parameterizations. In what follows we describe the outcomes of the results. 

\begin{itemize}
\item Using the SN+BAO+BBN+$H(z)$ data combination in a flat universe, AIC, BIC and DIC results indicate that SPad\'{e}-I as a dynamical DE scenario is the best model, even better than the $\Lambda$CDM model. In particular, DIC, being the most robust of all the information criteria, shows that we have ``significant support'' for SPad\'{e}-I compared to other Pad\'{e} parameterizations.

\item Using the SN+BAO+BBN+$H(z)$ data combination in nonflat universe, all the criteria indicate that $\Lambda$CDM is the best model. In this case $\Delta$AIC $<2$ indicates that there is ``significant support'' for SPad\'{e}-I. Furthermore, BIC shows that there is "no evidence" against SPad\'{e}-I. As $\Delta$DIC $<2$ for all of the parameterizations, this confirms that there is ``significant support'' for the present Pad\'{e} parameterizations.

\item Using SN+BAO+BBN+$H(z)$+CMB as the data combination in a flat universe, both AIC and BIC show that SPad\'{e}-I is the best model, which is followed by the concordance model. $\Delta$DIC values indicate that there is ``significant support'' for the Pad\'{e} parameterizations.

\item Using the SN+BAO+BBN+$H(z)$+CMB data combination in a nonflat universe and assuming all the criteria, we find SPad\'{e}-I as the best model, which is followed by $\Lambda$CDM. In this case, similar to the earlier case, we have $\Delta$DIC $<2$ for all models which means that there is ``significant support'' for all Pad\'{e} parameterizations. 
  
\end{itemize}

\section{Concluding remarks}
\label{sec-conclusion}

The nature of DE is unknown, at least based on the current observational evidence.  Although the $\Lambda$CDM cosmological model has been very successful in describing the present DE dominated era, there are significant hints in favor of a cosmological scenario beyond $\Lambda$CDM. Alternatively, cosmological models in which the EoS of DE is not equal to $-1$ have received significant attention.

In the present article we have investigated whether the current observational data indicate any deviation from $w =-1$. As the choice of the DE sector is not unique, we have adopted Pad\'{e} approximation, a well known approach to model several DE parameterizations. In particular, the Pad\'{e} approximation can recover some well known DE parameterizations, e.g. constant EoS in DE, CPL parametriztion~\cite{Chevallier:2000qy, Linder:2002et}, linear EoS in terms of the redshift $w(z) = w_0 + w_a z$~\cite{Cooray:1999da}, polynomial form of DE EoS $w (z) = \sum_{i}w_i z^i$~\cite{Astier:2000as}, and others. We have considered three specific dynamical EoS parameters for DE following Pad\'{e} parameterizations, labeled  Pad\'{e}-I, SPad\'{e}-I and Pad\'{e}-II, all of which recover $w =-1$ at the present epoch which
means they become nondynamical at the current epoch. More specifically, they recover the cosmological constant at the present epoch. 
On the other hand, 
we have also considered that our Universe could be either spatially flat or nonflat. This allows us to further understand whether the curvature of the Universe plays any crucial role in the determination of the deviation of DE from $w(a) = -1$.  The resulting scenarios have been constrained initially using two combined datasets, namely, SN+BAO+BBN+$H(z)$ and  SN+BAO+BBN+$H(z)$+CMB, and then we have added 1000 mock GWSS data sets matching the expected sensitivity of the Einstein Telescope to the combined probes of the real datasets. The motivation for adding the GWSS dataset is to examine how much the future GWSS dataset could offer the constraining power over the usual cosmological probes and in addition to that whether any hints in favor of the time-varying nature in the DE EoS could be obtained from the future GWSS.

Thus, considering all four combined datasets (with GW and without GW), the results are summarized in Tables~\ref{tab:flat} and \ref{tab:non-flat}. The graphical variations are also shown in Figs. \ref{fig:p1k0}, \ref{fig:spk0}, \ref{fig:p2k0} (for the spatially flat case) and in Figs. \ref{fig:p1}, \ref{fig:sp}, \ref{fig:p2} (for the nonflat case). 
We first focus on the curvature parameter of the Universe. According to the observational constraints on the spatial curvature of the Universe, although for some cases we observe discrete evidence for a non-null $\Omega_{k0}$, they are not enough to favor a nonflat universe. This observation is true for all the datasets (without or with GW). However, looking at the mean values of $\Omega_{k0}$ as observed from various Pad\'{e} parameterizations, it is evident that the inclusion of spatial curvature in the cosmological models should be examined in detail. Next we focus on the DE parameters that determine whether there is any evidence of dynamical DE and hence any deviation from $w(a) = -1$.

In a spatially flat universe, our observations are the following: {\bf (i)} a mild evidence of the dynamical DE EoS, and consequently, a deviation from $w(a) = -1$ within Pad\'{e}-I for both SN+BAO+BBN+$H(z)$ and  SN+BAO+BBN+$H(z)$+CMB is found, however, when the GW dataset is added to both these combined datasets, mild evidence of a dynamical DE EoS (and hence the deviation from $w(a) =-1$) still found for SN+BAO+BBN+$H(z)$+GW, but the resulting scenario is found to be consistent with $\Lambda$CDM for SN+BAO+BBN+$H(z)$+CMB+GW; {\bf (ii)} mild evidence of a dynamical DE within SPad\'{e}-I is suggested for both SN+BAO+BBN+$H(z)$ and SN+BAO+BBN+$H(z)$+GW, but the scenario is consistent with $\Lambda$CDM for SN+BAO+BBN+$H(z)$+CMB and SN+BAO+BBN+$H(z)$+CMB+GW; {\bf (iii)} significant evidence  of the dynamical EoS (at $6.4\sigma$) in DE within Pad\'{e}-II is found for SN+BAO+BBN+$H(z)$+CMB, and when GW is added to this combined dataset (i.e. for SN+BAO+BBN+H(z)+CMB+GW), this evidence is found at $4.7\sigma$. When CMB is not present in the combined datasets (i.e. for SN+BAO+BBN+$H(z)$ and SN+BAO+BBN+$H(z)$+GW), evidence is found at slightly more than $1\sigma$.  That means that the evidence of dynamical DE is strong when CMB is present in the combined datasets.

On the other hand, when the curvature of the Universe is allowed, then we find {\bf (i)} mild evidence of the dynamical EoS in DE within Pad\'{e}-I for both SN+BAO+BBN+$H(z)$ and SN+BAO+BBN+$H(z)$+CMB, but when the GW dataset is added to them the scenario is found to be consistent with $\Lambda$CDM for SN+BAO+BBN+$H(z)$+GW, but we get back the evidence of a dynamical DE and hence the deviation from $w(a) =-1$ for SN+BAO+BBN+$H(z)$+CMB+GW; {\bf (ii)} positive evidence of a dynamical EoS in DE and therefore the deviation from $w(a) =-1$ within SPad\'{e}-I for all the combined datasets (without and with GW); {\bf (iii)} significant evidence of the dynamical EoS  in DE within Pad\'{e}-II ($6.2\sigma$ for SN+BAO+BBN+$H(z)$ and $6.4\sigma$ for SN+BAO+BBN+$H(z)$+CMB). The addition of GW data slightly reduces the evidence ($3.8\sigma$ for SN+BAO+BBN+$H(z)$+GW and $5.1\sigma$ for SN+BAO+BBN+$H(z)$+CMB+GW), but the preference persists at a high confidence level. 
Moreover, one can further identify the changes in the dynamical nature in SPad\'{e}-I when the curvature of the Universe is allowed.

Having all the results in hand, one can clearly see that within the Pad\'{e}-II parameterization, a significant evidence of a  dynamical DE deviating from $w(a) =-1$ is found across all the datasets. Especially for the robust data combination SN+BAO+BBN+$H(z)$+CMB we have obtained $w(a) \neq-1$ at more than $6\sigma$. This outcome is very interesting and timely since according to recent DESI BAO data~\cite{DESI:2024uvr}, evidence for dynamical DE (even though such evidence is not as statistically strong as the Pad\'{e}-II parameterization)  has been found under the assumption of CPL parameterization \cite{DESI:2024mwx} and such evidence of dynamical DE also remains robust irrespective of other DE parameterizations~\cite{Giare:2024gpk}.   Although no such evidence of a nonzero spatial curvature of the Universe is suggested by any of the datasets, we noticed that the inclusion of the curvature  parameter can affect the DE constraints: for instance, within SPad\'{e}-I, evidence of dynamical DE  emerged in the presence of the curvature parameter.

Finally, we performed two distinct model comparison analyses, namely, the Bayesian evidence analysis  (see the results in Table~\ref{tab:BE}) and information criteria (AIC, BIC, DIC; see Table~\ref{tab:IC}).  According to the Bayesian evidence analysis, our conclusion is very clear: all the Pad\'{e} parameterizations are close to $\Lambda$CDM. Although from the statistical point of view, SPad\'{e}-I is mildly preference over $\Lambda$CDM for SN+BAO+BBN+H(z)+CMB (in the context of the flat universe), that is not so significant, and $\Lambda$CDM remains favored for all the remaining cases including datasets and models. Considering various information criteria, we see that according to AIC and BIC,  SPad\'{e}-I remains favored over Pad\'{e}-I  and Pad\'{e}-II. This might be due to fact that Pad\'{e}-I  and Pad\'{e}-II have one more parameter than SPad\'{e}-I. Additionally, we further notice that, except for the data combination SN+BAO+BBN+$H(z)$ in the nonflat case,   SPad\'{e}-I is preferred over $\Lambda$CDM as well. However, according to DIC, all three parameterizations  are close to $\Lambda$CDM. Essentially, both the model comparison analyses conclude that the present Pad\'{e} parameterizations lie very near the $\Lambda$CDM cosmological model.

These altogether suggest that the Pad\'{e} parameterizations are quite appealing for further investigations.  In fact, the models studied in this article are completely new, and so are the constraints on them. In particular, within the Pad\'{e}-II parameterization, the evidence of dynamical DE is significant. While such evidence is statistically robust, a more deterministic nature of this evidence can be expected in the presence of full CMB likelihoods from Planck and other experiments. 
Moreover,   with the emergence of other potential cosmological probes, such as DESI BAO DR2 \cite{DESI:2025zgx}, SN samples from distinct observational missions~\cite{Brout:2022vxf,DES:2024upw,Rubin:2023ovl}, the latest BBN data \cite{Schoneberg:2024ifp}, etc., more exciting news on the Pad\'{e} parameterizations can be expected.  Updated constraints on these Pad\'{e} parameterizations using the latest cosmological probes will be reported in a forthcoming work.

\section{Acknowledgments}
The authors sincerely thank the referee for many insightful comments which significantly improved the quality of the article.  
SP acknowledges the financial support from  the Department of Science and Technology (DST), Govt. of India under the Scheme  ``Fund for Improvement of S\&T Infrastructure (FIST)'' [File No. SR/FST/MS-I/2019/41]. WY was supported by the National Natural Science Foundation of China under Grants No. 12175096 and No. 11705079, and Liaoning Revitalization Talents Program under Grant no. XLYC1907098.  D.F.M. acknowledges support from the Research Council of Norway and UNINETT Sigma2 $-$ the National Infrastructure for High Performance Computing and Data Storage in Norway.

\bibliography{biblio}

\end{document}